\documentclass[final,twoside,11pt]{entics}

\usepackage{amssymb,amsmath,mathrsfs}

\usepackage{mathtools}
\usepackage{thmtools}
\usepackage{xspace}
\usepackage{dsfont}
\usepackage{tikz}
\usetikzlibrary{cd,fit,calc,positioning,arrows,automata,shapes}
\usepackage{ifthen}

\usepackage[noadjust]{cite}
\usepackage{rotating}
\usepackage{microtype}
\usepackage{graphicx}
\usepackage{amsbsy,eucal}  
\usepackage{dsfont}
\usepackage{xspace}

\renewcommand{\phi}{\varphi}

\newcommand{\lang}{\mathcal{L}}
\newcommand{\RR}{\mathcal{R}}
\usepackage{ifdraft}

\newcommand{\resetCurThmBraces}{%
\gdef\curThmBraceOpen{(}%
\gdef\curThmBraceClose{)}}
\resetCurThmBraces
\newcommand{\takeoutoveThmBraces}{%
\gdef\curThmBraceOpen{}%
\gdef\curThmBraceClose{}}
\resetCurThmBraces

\usepackage{etoolbox}
\patchcmd{\thmhead}{(#3)}{\curThmBraceOpen #3\curThmBraceClose }{}{}

\newcommand{\semantics}[1]{[\![ #1 ]\!]}

\usepackage[footnote,marginclue,nomargin]{fixme}

\newcommand{\one}{\mbox{\tt 1}}
\newcommand{\hash}{\mbox{\tt \#}}

\newcommand{\proves}{\vdash}

\newcommand{\eps}{\varepsilon}
\newcommand{\semalt}[1]{ {#1}^*}
\newcommand{\semaltparens}[1]{ {(#1)}^*}
\newcommand{\nf}{\mbox{\sf nf}}
\newcommand{\ddd}{\mbox{\tt d}}
\newcommand{\kkk}{\mbox{\tt k}}
\newcommand{\bbb}{\mbox{\tt b}}

\newcommand{\ccc}{\mbox{\tt c}}
\newcommand{\sss}{\mbox{\tt s}}
\newcommand{\eee}{\mbox{\tt e}} 
\newcommand{\www}{\mbox{\tt w}}
\newcommand{\uuu}{\mbox{\tt u}} 
\newcommand{\dc}{\mbox{$\ddd$-ct}}
\newcommand{\wc}{\mbox{$\www$-ct}}
\DeclareMathOperator {\app}{@}
\newcommand{\bfn}{b}
\newcommand{\bprimefn}{b'}
\newcommand{\soneone}{s^1_1}
\renewcommand{\o}{\circ}
\newcommand{\quadiff}{\quad \mbox{iff} \quad}
\newcommand{\Model}{\mathcal{M}}

\newcommand{\op}[1]{\operatorname{\mathsf{#1}}}
\newcommand{\id}{\op{id}}

\newcommand{\set}[1]{\{#1\}}

\renewcommand{\o}{\cdot}

\newcommand{\takeout}[1]{\empty}

\newcommand{\pair}[1]{\langle {#1} \rangle}

\newlength{\mathfrwidth}
\newsavebox{\mathfrbox}
\newenvironment{mathframe}
    {\begin{lrbox}{\mathfrbox}\begin{minipage}{\mathfrwidth}\begin{center}\begin{math}\displaystyle}
    {\end{math}\end{center}\end{minipage}\end{lrbox}\noindent\fbox{\usebox{\mathfrbox}}}

\newcommand{\diag}{\mbox{\emph{diag}}}

\newcommand{\Aomega}{A_{\omega}}

\input{enticsmacro.sty}

\usepackage[
    type={CC},
    modifier={by-nc-sa},
    version={3.0},
]{doclicense}

\makeatletter
\def\blfootnote{\xdef\@thefnmark{}\@footnotetext}
\makeatother

\def\conf{MFPS 2023} 	
\def\myconf{mfps39}
\volume{3}			
\def\volu{3}			
\def\papno{17}			
\def\lastname{Moss} 
\def\runtitle{Algebra of Self-Expression} 
\def\mydoi{12320}
\def\copynames{L. Moss}    

\begin{document}

\begin{frontmatter}
\title{Algebra of Self-Expression} 						
  \author{Lawrence S.~Moss\thanksref{myemail}\thanksref{Simons}}	
   \thanks[myemail]{Mathematics Department, Indiana University,		
Bloomington, USA 47401.  Email: \href{lmoss@indiana.edu} {\texttt{\normalshape
        lmoss@indiana.edu}}} 
            \thanks[Simons]{Supported by grant \#586136 from the Simons Foundation.} 



\begin{abstract} 
Typical arguments for results like Kleene's Second Recursion Theorem and the existence of self-writing computer programs bear the fingerprints of equational reasoning and combinatory logic. In fact, the connection of combinatory logic and computability theory is very old, and this paper extends this connection in new ways. In one direction, we counter the main trend in both computability theory and combinatory logic of heading straight to undecidability. Instead, this paper proposes using several very small equational logics to examine results in computability theory itself. These logics are decidable via term rewriting. We argue that they have something interesting to say about computability theory. They are closely related to fragments of combinatory logic which are decidable, and so this paper contributes to the study of such fragments. The paper has a few surprising results such as a classification of quine programs (programs which output themselves) in two decidable fragments. The classification goes via examination of normal forms in term rewriting systems, hence the title of the paper. The classification is an explanation of why all quine programs (in any language) are ``pretty much the same, except for inessential details."
In addition, we study the relational structure whose objects are the programs 
with the relation $p\to q$ (read ``$p$ expresses $q$'') if program $p$ when run on nothing outputs $q$.
\end{abstract}
\end{frontmatter}


 \maketitle

\section{Introduction}

The modest point made in this paper is that several results in computability theory 
call on equational reasoning of a very simple sort.  
We are thinking of the equations seen in the $s^m_n$-Theorem or the universal program,
and of even more interest here are equations pertaining to a ``diagonal operator'' that 
one can see in proofs of Kleene's Second Recursion Theorem.
We formulate several systems
of equations 
(as in abstract data types or equational logic)
which can then be studied using term rewriting theory.   The particular
systems of equations which we study turn out to be terminating and confluent,
and thus the initial algebras are computable.   They are also quite close to 
fragments of combinatory logic, but the fact that they have easily-described 
computable models means that they are much weaker than combinatory
logic as one usually sees it.   This  loss of  expressive power has a compensation:
we can study the normal forms in our rewriting systems and prove a few results
which are unexpected, such as a characterization of quine programs in the
language of the diagonal operator and program sequencing.

\subsection{Kleene's Second Recursion Theorem: two proofs}

Let us begin with one of the cornerstone results in computability theory, 
Kleene's Second Recursion Theorem~\cite{Kleene38}.
This result is more often called \emph{the Recursion Theorem}.
We are not interested in its many applications.
(For some of those, especially in classical recursion theory and descriptive set theory, 
see Moschovakis~\cite{Moschovakis}.
For other applications, this time in connection with 
computer viruses, see Bonfante et al~\cite{BonfanteKM07}.)
Instead, we are only interested here in the proof.
We are going to formulate this in terms of \emph{programs} in some unspecified language,
whereas the classical result uses natural numbers as \emph{indices} of \emph{computable functions}.
In the classical setting, a number $e$ is an index of an $n$-place partial function $\phi^{n}_{e}(x_1, \ldots, x_n)$.
In our setting, or in any setting involving nested use of this notation, it is more readable to write this
as $\semantics{e}^n(x_1,\ldots, x_n)$.  Indeed, we also usually drop
 the superscript $n$; it will always be clear from the context.
We are going to state the Recursion Theorem for $n =2$ for simplicity.

\takeout{
My plan here is to present the two proofs of the Second Recursion Theorem
and also mention the points that Neil Jones' paper makes about those proofs.
This is motivation for the equational presentation in Section~\ref{section-equation-d}.
}

\begin{theorem}\label{K2R}
Let 
$p$  be a program,  consider as a function   $\semantics{p}$ of two arguments.
There is a program $q^*$ so that for all $r$,
$$\semantics{q^*}(r)  \simeq\semantics{p}(q^*, r).$$
\end{theorem}

\begin{proof} Here is the original proof due to Kleene. 
 Recall the program $s^1_1$; it has the property that
\begin{equation}\label{eqs11}
\semantics{\semantics{s^1_1}(x,y)}(z) = \semantics{x}(y,z).
\end{equation}
Fix $p$ and then modify it
to get $p'$ with the property that for all $x,r$,
\[ \semantics{p'}(x,r) = \semantics{p}(\semantics{s^1_1}(x,x), r) . \]
Then take
 $q^*$ to be $\semantics{s^1_1}(p',p')$.  
 To see  that $q^*$
defined this way has the property that we want, we calculate:
\[
\semantics{q^*}(r) = \semantics{\semantics{s^1_1}(p',p')}(r) 
= \semantics{p'}(p',r) 
=   \semantics{p}(\semantics{s^1_1}(p',p'), r) 
= \semantics{p}(q^*, r).
\]

\takeout{
\[
\begin{array}{lcl}
\semantics{q^*}(r) & = & \semantics{\semantics{s^1_1}(p',p')}(r) \\
& = & \semantics{p'}(p',r) \\
& = &   \semantics{p}(\semantics{s^1_1}(p',p'), r) \\
& = &  \semantics{p}(q^*, r)
\end{array}
\]
}
Here is a second proof, from~\cite{Moss06}.
Modify $p$ to get  $\hat{q}$ so that for all $x$ and $r$:
\[\semantics{\semantics{\hat{q}}(x)}(r)
\ \simeq \
 \semantics{p}(\semantics{x}(x),r).
\]
Then we can take $x$ to be $\hat{q}$ itself,
and we would have 
that for all $r$,
\[\semantics{\semantics{\hat{q}}(\hat{q})}(r)
\ \simeq \
 \semantics{p}(\semantics{\hat{q}}(\hat{q}),r).
\] 
Thus when we take $q^*$ to be  $\semantics{\hat{q}}(\hat{q})$,
we would have 
$\semantics{q^*}(r)
\simeq 
 \semantics{p}(q^*,r)$.
\end{proof}

\begin{narrower}
There are several questions which we want to ask about the two proofs.
Are they the same?   If not, what is common to them?  Why should anyone care?
\end{narrower}

The two proofs are of course not the same.  This matter was 
 studied by Jones~\cite{Jones13} in connection with in-practice computational systems that
 make use of the Recursion Theorem.  It turns out that the two different constructions
  are correlated with performance differences in a 
 certain algorithm mentioned in~\cite{Jones13}.   Be that as it may, there is evidently something similar
 about the two proofs.
 They both 
 employ what we called ``modification''.  This is a very general technique and we shall see a version of it
 under the name of \emph{combinatory completeness}.   
In the original proof, Kleene is using (\ref{eqs11}) and a bit of \emph{equational reasoning}.
 
 This paper is not about the Recursion Theorem per se but about equational reasoning 
 in both of these proofs.  Indeed, we are interested in equational logics which are ``small'' enough to
 be decidable and yet ``big'' enough to say something interesting about fundamental results in computability 
 theory.   We ultimately would like to return to the  Recursion Theorem and if possible to classify 
 its equational proofs.  What we have done in this paper is to take a step towards this goal by formulating
  an equational logic related to the diagonalization operation $p\mapsto\semantics{p}(p)$ that we see in the
  second proof of the Recursion Theorem.    Our main results concern two equational systems related to
  this operation.
 The reason why we are not yet able to formulate a logic that covers the first proof as well comes from the
 fact that this proof is more about substitution than about diagonalization.   To make this point clearer,
 we digress to study quine programs.

\subsection{Some problems about program ``expression''}

A \emph{quine program} is a program $p$ such that when $p$ is run ``without input'',
the output is $p$.   We have drawn attention to the ``without input'' point.
To formalize this as we have done above, we would say that $\semantics{p}^0(\ ) \simeq p$.
When we think of programs via natural number indices, this would seem to be a natural move.
However,
when we think of them as programs which operate on ($k$-tuples of) words for some $k$,
 we could equally well say 
 a quine program is a program $p$ such that $\semantics{p}^1(\eps) \simeq p$, where $\eps$ is the empty word.

Perhaps the easiest way to construct such a program is to use the Recursion Theorem, but not exactly in the form
which we saw in Theorem~\ref{K2R}.  We want a slightly different version:
Let 
$p$  be a program, and consider the function  $\semantics{p}^1$.
There is a program $q^*$ so that 
$\semantics{q^*}(\ )  \simeq\semantics{p}(q^*)$.
Then we would take $p$ to be any program such that $\semantics{p}$ is the identity function.
For this $p$, the promised program $q^*$is a quine program.

One classical construction of a quine program
in a sense bypasses the much more general (and thus better) construction offered by the Recursion Theorem.
It  can be summarized as follows.
Let $\diag$ be a program such that  for all $x$, $\semantics{\semantics{\diag}(x)}(\ ) \simeq \semantics{x}(x)$.
Then $\semantics{\semantics{\diag}(\diag)}( \ )  \simeq \semantics{\diag}(\diag)$.
So $\semantics{\diag}(\diag)$ is a quine program.
For an example, here is the quine in Lisp:
\[
\mbox{\tt ((lambda (x) (list x (list 'quote x))) '(lambda (x) (list x (list 'quote x))))}
\]
So here $\diag$ would be the program $\mbox{\tt ((lambda (x) (list x (list 'quote x)))}$.
However, in other languages, the most direct quine program would not be obtained in this way.
For example, looking at quines in Python on the internet, I have not seen any that seem to
come from a $\diag$ program in this way.  Instead, they rely on string substitution.
I have in mind examples like
\[
\begin{array}{l}
\texttt{var = "print('var = ', repr(var), '\textbackslash\textbackslash neval(var)')"}\\
\mbox{\tt eval(var)}\\
\end{array}
\]
Getting an account of all the possible quine programs using substitution is beyond the scope of this paper.
What we do is to construct an equational logic related to the diagonalization operator and prove results about that.

\subsection{Program expression}
The quine program
may be put into a larger context.  
Marvin Minsky wrote ``It is generally recognized that 
the greatest advances  in modern computers
came through the notion that
programs could be kept in the same memory  with `data,'  
and that programs  could operate on other programs, 
or on themselves, 
as though they were data.''
Let us take seriously the idea that programs can \emph{output} other programs as well.
 For programs $p$ and $q$,  write
\begin{equation}\label{express}
p \to q \quadiff \semantics{p}(\ ) \simeq q
\end{equation}
We read this as $p$ ``expresses'' $q$.  
That is, running $p$ on ``nothing'' outputs $q$.   
 If we take seriously Minsky's point, it would seem that the study of the
 graph $(\mbox{\emph{Programs}},\to)$ would be of interest in computer science.
 A quine program is one which expresses itself.  Are there pairs of different programs 
 which express each other, or about cycles of other lengths?   Are there infinite sequences
 $p_0 \to p_1 \to p_2 \to \cdots $ of distinct programs?
The answers to both of these questions is Yes.  There are cycles of all lengths
 and infinite sequences.  The proofs (at least the only proofs which I knew before working on this topic)
 used the Recursion Theorem, or something close.  They were not ``equational.''  So the purpose of
 this paper is to ask about equational proofs of the main facts of the program-expression graph.

\subsection{Plan for the paper}

We are concerned with the equational reasoning that goes on in the basic parts of computability.
So the main proposal in the paper is to formulate an equational logic whose operations include
sequencing of instructions (written with the symbol $+$) as well as application of one program to another
(written $\app$, and sometimes elided).   The constants are the empty program $\eee$,
the program $\soneone$, and the universal programs $\uuu_0$ and $\uuu_1$.
In addition, we are particularly interested in two constants which are not part of the standard 
presentations of computability theory.  One is $\ddd$, related to the diagonal map $x\mapsto \semantics{x}(x)$.
(But our intended interpretation of $\ddd$ is not quite this map.) The other is $\www$, intended to be
$x\mapsto$ ``the instructions to write $x$''.    The set of equations pertinent to all of this is in Figure~\ref{fig-3}.

Our plan is to study this equational logic in fragments.  
Following a short look at the background in Section~\ref{section-background},
we begin with one of the two main fragments in the paper: 
the one with $\eee$, $\ddd$, $+$ and $\app$;
this is studied in Section~\ref{section-equation-d}.
In Section~\ref{section-equation-w}, we add $\www$.  There is no section devoted only to 
 $\eee$, $\www$ (rather than $\ddd$), $+$ and $\app$: although this fragment is natural to consider, it is too weak for the purposes
 of this paper.   We have a few minor remarks about the larger fragments and some other topics in 
 Section~\ref{section-large}.  Two discussions have been moved to the Appendix.

\begin{figure}[t]
\begin{mathframe}
\begin{array}{lcl}
x + \eee  & = & x \\
\eee + x & = & x \\
(x + y) + z & = &  x + (y+z) \\
(x + y ) \app z &= & y \app (x \app z) \\
\end{array}
\qquad\quad
\begin{array}{lcl}
\ddd\app x & = & (\www \app x) + x \\
\eee\app x & = & x \\
\www \app (x +  y)& = & (\www \app x ) +  (\www \app y)\\
\www\app x\app y & = & y + x\\
\end{array}
\qquad\quad
\begin{array}{lcl}
\www\app \eee & = & \eee\\
\sss^1_1\app x \app y\app z  & = & x\app y \app z \\
\uuu_0 \app x & = & x \app \eee\\
\uuu_1 \app x \app y & = & x \app y\\
\end{array}
\end{mathframe}
\caption{The full set $E$ of equations used in this paper.
\label{fig-3}}
\end{figure}

\section{Background}
\label{section-background}

The background needed to read this paper is very minimal. 
The main required notions are those from equational logic and term rewriting
theory.  We review this shortly.   It would be good to have seen combinatory logic (CL), but this is
mainly to compare what we are doing to CL.   Except for that material, the paper should
be readable even for someone who has never seen CL.

\subsection{Equational logic, term rewriting,  and computable algebras}

Let $E$ be a set of equations.
We write $t\equiv u$ if $E\proves t=u$ using $E$ and the inference rules of equational logic.
These rules are the reflexive, symmetric, and transitive properties of equality, substitution,
and congruence for all function symbols.
We assume familiarity with the definitions and basic results mentioned just above.
Here are the results which we shall be using.

\begin{theorem} [\cite{MeseguerGoguen}, Theorem 51]
Suppose that $E$ is a finite set of equations
which can be \emph{oriented to give a term rewrite system}
which is \emph{terminating} and \emph{confluent}.
Then every term $t$ has a unique \emph{normal form} $\nf(t)$.
Moreover, we have
\[\mbox{$t \equiv u$ \quadiff $\nf(t) = \nf(u)$.}\]
The set of normal forms is the \emph{initial algebra of $E$}
with a canonical structure: 
\[ f(\nf(t_1), \ldots, \nf(t_n)) = \nf (f(t_1, \ldots t_n)).\]
 This algebra 
 is a \emph{computable algebra}.
 \label{theorem-MG}
\end{theorem}

The assumption in this last result that $E$ can be oriented to give a terminating
and confluent term rewriting system is rather strong.   In general, it fails.

\begin{theorem}[Perkins~\cite{Perkins}]
The question, given a finite set $E$ of equations 
and an equation $t=u$,
of whether or not $t \equiv_E u$, is undecidable.
\end{theorem}

One concrete setting where this applies is to combinatory logic, reviewed just below.

\subsection{Combinatory algebras}
\label{sec-ca}
For background on combinatory logic, see~\cite{bimbo}.

\begin{defn} A \emph{combinatory algebra (CA)} is a tuple $(D,\cdot,\sss,\kkk)$ such that $D$ is a set,
$\cdot$ is a (total) binary operation on $D$, $\sss$ and $\kkk$ are elements of $D$,
and 
\[\begin{array}{lcl}
\kkk \cdot x \cdot y  & = & x \\
\sss\cdot x \cdot y \cdot z & = & x \cdot z \cdot (y \cdot z) \\
\end{array}
\]
In the equations above, we followed the standard notation of omitting parentheses.
For example  $\sss\cdot x \cdot y \cdot z $ is really $(((\sss\cdot x) \cdot y )\cdot z) $,
and $ x \cdot z \cdot (y \cdot z)$ is really  $(x \cdot z) \cdot (y \cdot z)$.
In addition, one usually omits the $\cdot$ entirely and writes, for example, $\sss x y z = x z (yz)$.

\end{defn}

\begin{theorem}[Combinatory Completeness]
For every term $t(x_1, \ldots, x_n)$
 in variables $x_1$, $x_2$, $\ldots$, $x_n$ there is an element $t^*\in D$
such that $t^*\cdot d_1 \cdot d_2 \cdot\cdots \cdot d_n = t(d_1,\ldots,d_n)$ for all $d_1$, $\ldots$, $d_n\in D$.
\end{theorem}

\begin{example}
With $t$ the term $x(yz)$, we can take $\bbb = \sss\cdot (\kkk \cdot \sss)\cdot \kkk$.
This is because 
\[
\sss (\kkk  \sss) \kkk  x  y  z 
= (\kkk\sss) x (\kkk y) z = \sss (\kkk x) y z = \kkk x z (y z) = x (yz).
\]
Another example: let $t$ be the term $y(x(z))$.
Let $\ccc = 
\sss(
\bbb\bbb\sss)(\kkk\kkk)$, and 
 let $\bbb' = \ccc\bbb$.  Then one can show
 that $\bbb'$ has the required property: $\bbb'x y z = y(xz)$.
The details of why this works are not important here; the main thing is the combinatory completeness result itself.
\end{example}

Here is an outline of the general method for \emph{solving fixed point equations} in CL, using combinatory completeness.
Consider a function $f(x)$ such as $f(x) = x\cdot e$.   We show that  there is  a term $x^*$ so that $f(x^*) \equiv x^*$.
For this,
consider the term $f(x\cdot x)$.  By combinatory completeness, let $t^*$ be such that 
for all $x$, $t^*\cdot  x \equiv f(x\cdot x)$.
Let $x^* = t^*\cdot t^*$.
Then 
\[
x^* =  t^*\cdot t^* \equiv f(t^* \cdot t^*) = f(x^*).
\]
The second proof of Kleene's Second Recursion Theorem is basically 
an application of this general method.   In our second proof of Theorem~\ref{K2R},
 one starts with $\semantics{p}(x,r)$ and then considers
$\semantics{p}(\semantics{x}(x)r)$.  Now $\semantics{x}(x)$ is quite closely related to $x\cdot x$.
Indeed, one would like to take the natural numbers (indices of partial computable functions)
and then construct a combinatory algebra using $x\cdot y = \phi_x(y) = \semantics{x}(y)$
together with chosen indices to interpret $\sss$ and $\kkk$.  The problem with this is that 
combinatory algebras are structures in the sense of universal algebra and thus must interpret
their symbols by total functions.  So this idea does not work out -- but it shows the connection between
fixed points in CL and this proof of the Recursion Theorem.  

Here are some further comments related to this last point.  First, this issue of total and partial interpretations
will return in this paper in interesting ways.  Second, we mention other work related to combinatory logic which
and category theory which might well be connected to our subject.   These are the recent paper by Roberts~\cite{Roberts}
on general settings for diagonalization arguments,
and the older paper by Di Paola and Heller~\cite{DiPaola} which offers a category-theoretic 
treatment of partiality in computability theory.

\takeout{
\begin{theorem} 
Every computable combinatory algebra is trivial.
\end{theorem}
}

\takeout{
\subsection{Partial combinatory algebras}
\label{section-PCA}

My plan is to present the Kleene PCA here since this gives a model of the equational theory
which could be understood by a person who knows computability theory but not $\one\hash$.
}

\subsection{Related work}

To the best of our knowledge, there are 
no studies of the equational systems proposed in this paper.
There are decidable fragments of combinatory logic;
for example, see~\cite{Waldmann,Statman}.
Our proposal is (with minor differences)
another decidable fragment of CL.   However, our work is different for
a number of reasons.  First and foremost, our motivation comes from the equational analysis of
basic results in computability, not from CL itself.   Second, the particular equational systems here
are different from what we have seen in the literature, and even small differences in equational systems
sometimes result in very  different technical developments.

\section{Equational logic of diagonalization}
\label{section-equation-d}

We present an equational logic in the style of combinatory logic (Section~\ref{sec-ca}).
We take constants $\ddd$ and $\eee$,
and also two binary function symbols, $\app$ (for application) and $+$ (for concatenation).
In the language of combinatory logic,  $t\app u$ is just a variant symbol for $t\cdot u$ (usually written $tu$) 
and $t+u$ is a variant notation for $\bbb \cdot t\cdot u$.
We are interested in the set $E$ of equations shown in Figure~\ref{fig-1}.
In later sections, we add
constant symbols for other basic operations in computability theory and also add equations
 to $E$.
 
\begin{figure}[t]
\begin{mathframe}
\begin{array}{lcl}
x + \eee  & = & x \\
\eee + x & = & x \\
(x + y) + z & = &  x + (y+z) \\
(x + y ) \app z &= & y \app (x \app z) \\
\end{array}
\qquad
\begin{array}{lcl}
(\ddd \app x) \app y & = & x \app (y + x)\\
\eee \app x & =& x \\
\ddd \app\eee & = & \eee\\
\\
\end{array}
\end{mathframe}
\caption{The set $E$ of equations for the language studied in Section~\ref{section-equation-d}.  The
signature has constants
$\ddd$ and $\eee$, and function symbols $+$ and $\app$.\label{fig-1}}

\end{figure}


We speak of $\app$-terms and $+$-terms, and these are defined in the evident way
by looking at the top-most symbol.  For example $\ddd\app (\eee + \ddd)$ is an $\app$-term,
and $(\eee \app \ddd) + \ddd$ is a $+$-term.

\subsection{The idea behind the equations, in plain terms}
\label{plainterms}

Recall the quote from Minsky emphasizing how important it was to see that ``programs could operate on other programs, 
or on themselves, 
as though they were data.''
Let us think of  $\app$ as application of one program to another, and $+$
as  sequentialization, and let us restrict attention to languages  $\lang$ where $\app$ and $+$ are first-class constructs.
At first glance one might expect the equations in this paper to hold for $\lang$.
  But this cannot literally be right,
since the application construct is partial.    Replacing $=$ by the weaker $\simeq$, we would expect the
equations to all hold.  However, we want to work in standard equational logic and so we want to think about
total interpretations.  Thus, we would like 
an interpretation of the initial algebra of our equations into $\lang$
\[
{}^*\colon T_{\Sigma}/\!  E \to \lang
\]
We are not going to exhibit this map ${}^*$ for any real programming languages
(but see our earlier example of a quine in Lisp).  
Instead, we look to a stylized version of a programming language.  So we are looking at pseudocode and showing how one
can interpret the symbols $+$, $\app$, $\ddd$, and $\eee$ in such a way as to satisfy the equations.

Let us consider  a set $P$ of ``programs" written in English.\footnote{Readers unhappy with this may omit this section
and jump to Section~\ref{section-pickup}.  Very little in the rest of this paper depends on the interpretation of our
terms into programs in language $P$.}
These are  sequences of \emph{instructions}.
We only want instructions of a very simple form, including basic imperative verbs like \emph{sit} and \emph{stand}.
These basic programs should be intransitive verbs (no object).  For a transitive verb like \emph{write},
we want the ability to take an input word $w$ (in the ``register'', see below) and 
replaces the register contents with a program which 
would write $w$.
 
Let us spell out this idea in a little more detail, in effect defining $P$ 
(somewhat informally) and also giving its operational semantics.

For example, $P$ should contain the program \emph{sit}.  
Executing this program causes the person doing so to sit.
$P$ also should contain the program \emph{write ``sit''}.
Executing this program causes the person doing so to write the word ``sit''.
This writing should go on a \emph{register}.  We might think of this register 
as a blank page to start.  But even if the register were not blank, it would make
sense to write the word ``sit'' into the register.   For this, we write onto the end of the register.
This would be the opposite end from what we read from.
$P$ should be closed under concatenation.  For this, let us use a comma.
So $P$ should have a program \emph{stand, write  ``sit''}.

Now we can 
endow $P$ with the structure of a $\Sigma$-algebra which satisfies the equations in Figure~\ref{fig-1}.
We interpret $\eee$ by the empty program, and we take $+$ to be the operation taking programs $p$ and $q$
to be ``$p$, $q$''
when $p$ and $q$ are non-empty, and otherwise is just the ordinary concatenation of $p$ and $q$.
  Notice the comma here in this first case. 
Then the first three equations in Figure~\ref{fig-1} hold.

 We add the comma mainly because it looks better to do so.  But we could resort to other devices: we could
 start instructions with upper-case letters, for example.

Let  $\ddd^* = \mbox{\emph{diag}}$
and $\www^* = \mbox{\emph{write}}$,
 the programs 
given below:
 \renewcommand{\arraystretch}{.3}
\begin{equation}\label{notfunny}
\begin{array}{l@{\qquad}l}
 \mbox{\emph{write}}: &
\mbox{\emph{write the instructions to write what is in the register
}}\\
 \mbox{\emph{diag}}: &
\mbox{\emph{write the instructions to write what is in the register in front of it}}
\end{array}
\end{equation}
For example, executing the program $\mbox{\emph{write}}$
when \mbox{\emph{diag}} is in the register results in the register having
\[
\begin{array}{l}
\mbox{\emph{write ``write the instructions to write what is in the register in front of it''}}
\end{array}
\]
Executing the program $diag$ when the register contains the program \emph{sit} 
results in the register having the program \emph{write ``sit'', sit}.   Then executing this last program
with a word $q$ in the register
would cause the one doing so to write ``sit'  at the end of the register (so the register will contain \emph{$q$, sit}),
and then to sit.

 \renewcommand{\arraystretch}{1.5}

Further, let us run $diag$ when the register contains \emph{stand, sit}. 
We would get 
\[\emph{write ``stand, sit'', stand, sit}\]
If we now run this program when the register contains $q$, then the register would contain \emph{$q$, stand, sit}
and the executor would stand and then sit.

Let $(t \app u)^*$ be the result of running $t^*$
with $u^*$ in the register. 
(We also describe this by saying that we \emph{apply $t^*$ to $u^*$}.)
 That is, we interpret the binary function symbol $\app$ 
by the function $\mbox{\emph{app}}\colon P^2\to P$ given by
\[
\mbox{\emph{app}}(p,q) = \emph{the program in the register after we apply $p$ to $q$}
\]
With a sufficiently rich base language, this operation will be (merely) partial.  
We thus are not claiming that our language $P$ is itself a model of the equations under consideration.
We do claim that the interpretation function \mbox{\emph{app}} is total on the image of ${}^*$.
We postpone the discussion of this to Section~\ref{section-postpone}.

One of our equations is $(x+y)\app z = y\app (x\app z)$.
Suppose that $x$ is \emph{write ``die''},  $y$ is \emph{write ``hop''}, and $z$ is the empty word.
Then $x+y$ is \emph{write ``die'', write ``hop''}, and  $(x+y)\app z$ would give \emph{die, hop} in the register.
On the other side, $x\app z$ would give \emph{die} in the register, and so executing \emph{write ``hop''} on this
would give  \emph{die, hop}.

Another of our equations is $(\ddd\app x)\app y=  x \app (y + x)$.
As a way of verifying it by example, take $x$ to be \emph{hop} and $y$ to be \emph{stand}.
Then $\ddd\app x$ is \emph{write ``hop'', hop}.
Executing this with the register containing $y$ would give \emph{stand, hop} in the register, and the executor hops.
This is the interpretation of $(\ddd\app x)\app y$.
On the other side, $y+x$ is \emph{stand, hop}, and executing $x$ when this is in the register does not change the 
register, but the executor hops.

The last two equations are
$\eee\app x = x$  and $\ddd\app \eee = \eee$.
Since we take $\eee$ to be the empty program, these seem sensible.   But if we take $\eee$ to be some other
``do nothing'' program (think of \emph{skip}), then it might well be the case that $\ddd\app \eee\neq \eee$.
Indeed, this last equation $\ddd\app \eee = \eee$ might not hold ``on the nose.''  It might hold
``up to extensionality''.  That is, if we add a rule of extensionality to equational logic, inferring $x = y$ from $xz = yz$, then 
$\ddd\app \eee = \eee$ follows as a consequence.

\subsection{The quine program in this example}

 \renewcommand{\arraystretch}{.3}
 
Recall that we are defining a function ${}^*\colon T_{\Sigma}/E \to P$. 
We shall see in Example~\ref{ex-quine} that $(\ddd\app \ddd)\app \eee =  \ddd\app \ddd$.
 This suggests that $(\ddd\app\ddd)^*$ is a quine program: when run on an empty register,
 it ``expresses'' (outputs in the register) itself.
 To check this, note that 
 $(\ddd\app\ddd)^*$ is 
 \[\begin{array}{l}
 \mbox{\emph{write ``write the instructions to write what you see in front of it''}},
 \mbox{\emph{write the instructions}}\\
 \mbox{\emph{to write what you see in front of it}}
 \end{array}
  \]
 Then the reader can check that executing this program on an empty register
 outputs itself.
  \renewcommand{\arraystretch}{1.5}
 
 \subsubsection{A comment on the equations}
We \emph{don't want} the
  \emph{directly self-referential} program \emph{``print this sentence''}
 because the issue under discussion is whether we can achieve 
 the effect
of this, but  
 with much less powerful means.

\takeout{
We'll allow a small amount of quotation.

print the instructions to print what you see before it
Here "what you see" and "it" refer to the input, and "before" means "to the left of".
Applying this English program to the English program print me would give
print ``p'' print ``r'' print ``i'' print ``n'' print ``t'' print `` '' print "m" print ``e'' print me
Let's apply this English version of diag to itself.
We get the longer program below. It is the English version of self.


\begin{tabular}{l}
print ``p'', print ``r'', print ``i'', print ``n'', print ``t'', print `` '', print ``t'',  \\
print ``h'', print ``e'', print `` '', print ``i'', print ``n'', print ``s'', print ``t'', \\
print ``r'', print ``u'', print ``c'', print ``t'' , print ``i'', print ``o'', print ``n'',\\
 print ``s'', print `` '', print ``t',' print ``o'', print `` '', print ``p'' , print ``r'', \\
 print ``i'', print ``n'', print ``t'', print `` '', print ``w'', print ``h'', print ``a'',\\
 print ``t'' , print `` '', print ``y'', print ``o'', print ``u'', print `` '', print ``s'',  \\
print ``e'', print ``e'', print `` ''  , print ``b'', print ``e'', print ``f'', print ``o'', \\
print ``r'', print ``e'', print `` '', print ``i'', print ``t'', print the instructions to\\
print what you see before it
\end{tabular}


Andd this same explanation applies to the $\one\hash$ diag that we've already seen.

Let $diag$ be the program
\begin{equation}\label{notfunny}
\mbox{  print the instructions to print what you see in front of it}
\end{equation}

Running $diag$ on ``hide your face'' gives
 
\begin{tabular}{l}
print ``hide your face'', 
hide 
your face
\end{tabular}

Running $diag$ on ``die'' gives
\begin{tabular}{l}
print ``die'', 
die
\end{tabular}

Running ``diag'' on (\ref{notfunny}) gives
\begin{tabular}{l}
print ``print the instructions to print what you see in front of it'',\\
print the instructions to print what you see in front of it
\end{tabular}

Again, this is the program $diag$ applied to itself.
Let's execute this.
The first step gives

\begin{tabular}{l}
print the instructions to print what you see in front of it
\end{tabular}

and now the second step gives

We are interested in ``programs" of English (that is, sequences of instructions).

\

We want instructions of a very simple form, including 

\begin{itemize}
\item[$\triangleright$] instructions to print
 various characters, 
 
 \item[$\triangleright$] instructions to print instructions which would print
 various characters, 
 
\item[$\triangleright$] instructions which accept text as an input

\item[$\triangleright$] the ability to move text around

\item[$\triangleright$] the ability to sequence instructions or even programs

 \end{itemize}
 
 \
 
 But we \emph{don't want} the
 
  \emph{directly self-referential ``print this sentence''}
 
 because the issue under discussion is whether we can achieve 
 
 the effect
of this, but  
 with much less powerful means.

We'll allow a small amount of quotation.

print the instructions to print what you see before it

Here "what you see" and "it" refer to the input, and "before" means "to the left of".
Applying this English program to the English program print me would give

print ``p'' print ``r'' print ``i'' print ``n'' print ``t'' print `` '' print "m" print ``e'' print me

Let's apply this English version of diag to itself.
We get the longer program below. It is the English version of self.

And this same explanation applies to the $\one\hash$ diag that we've already seen.

Let $diag$ be the program
\begin{equation}\label{notfunny}
\mbox{  print the instructions to print what you see in front of it}
\end{equation}

Running $diag$ on ``hide your face'' gives
\begin{tabular}{l}
print ``hide your face'', 
hide 
your face
\end{tabular}

Running $diag$ on ``die'' gives
\begin{tabular}{l}
print ``die'', 
die
\end{tabular}

Running ``diag'' on (\ref{notfunny}) gives
\begin{tabular}{l}
print ``print the instructions to print what you see in front of it'',\\
print the instructions to print what you see in front of it
\end{tabular}

\

Again, this is the program $diag$ applied to itself.
Let's execute this.

The first step gives
\begin{tabular}{l}
print the instructions to print what you see in front of it
\end{tabular}

and now the second step gives

\begin{tabular}{l}
print ``print the instructions to print what you see in front of it'',\\
print the instructions to print what you see in front of it
\end{tabular}

The point is that executing the program at the top 

produced that very program.

Running $diag$ on its own text (see equation (1)) gives

\begin{tabular}{l}
print ``p'', print ``r'', print ``i'', print ``n'', print ``t'', print `` '', print ``t'', print ``h'', \\
print ``e'', print `` '', print ``i'', print ``n'', print ``s'', print ``t'', print ``r'', print ``u'',\\
print ``c'', print ``t'' , print ``i'', print ``o'', print ``n'', print ``s'', print `` '', print ``t','\\
print ``o'', print `` '', print ``p'' , print ``r'', print ``i'', print ``n'', print ``t'', print `` '', \\
print ``w'', print ``h'', print ``a'', print ``t'' , print `` '', print ``y'', print ``o'', print ``u'',\\
print `` '', print ``s'', print ``e'', print ``e'', print `` ''  , print ``b'', print ``e'', print ``f'', \\
print ``o'', print ``r'', print ``e'', print `` '', print ``i'', print ``t'',\\
print the instructions to print what you see before it
\end{tabular}

which as we have seen is a self-writing program.

What the observation about the supposed uniqueness of quines amounts to.

Generally speaking, self-writing programs are obtained 
by applying a program like $diag$ to itself.
Obviously, $diag$ can be written in slightly different ways,
and so quines are for sure not unique.
But suppose someone gives you \emph{only} 
`diag''
and some ``general constructions'' like
applying a program to an argument,
 the ability to sequence programs

We are interested in ``programs" of English (that is, sequences of instructions).

\

We want instructions of a very simple form, including 
instructions to print
 various characters, 
  instructions to print instructions which would print
 various characters, 
  instructions which accept text as an input
the ability to move text around
the ability to sequence instructions or even programs

 But we \emph{don't want} the
  \emph{directly self-referential ``print this sentence''}
 because the issue under discussion is whether we can achieve 
 the effect
of this, but  
 with much less powerful means.

Running $diag$ on ``hide your face'' gives
\begin{tabular}{l}
print ``hide your face'', 

hide your face
\end{tabular}

Running $diag$ on ``die'' gives
\begin{tabular}{l}
print ``die'', 
die
\end{tabular}

\begin{tabular}{l}
print ``print the instructions to print what you see in front of it'',\\
print the instructions to print what you see in front of it
\end{tabular}

Again, this is the program $diag$ applied to itself.
Let's execute this.
The first step gives
 
\begin{tabular}{l}
print the instructions to print what you see in front of it
\end{tabular}

and now the second step gives

\begin{tabular}{l}
print ``print the instructions to print what you see in front of it'',\\
print the instructions to print what you see in front of it
\end{tabular}

The point is that executing the program at the top 
produced that very program.

Running $diag$ on its own text (see equation (xxx)) gives
Generally speaking, self-writing programs are obtained 
by applying a program like $diag$ to itself.
Obviously, $diag$ can be written in slightly different ways,
and so quines are for sure not unique.
But suppose someone gives you \emph{only} 
the program $diag$
and some ``general constructions'' like
 applying a program to an argument,
 and the ability to sequence programs

The point is that executing the program at the top 
produced that very program.

Generally speaking, self-writing programs are obtained 
by applying a program like $diag$ to itself.
Obviously, $diag$ can be written in slightly different ways,
and so quines are for sure not unique.

So now we can ask the first question in this paper:
Is there a unique self-writing program?

\subsection{A model, preliminary discussion}
\label{section-model-preliminary}

Here is a model of the set of $E$ equations in Figure~\ref{fig-1}.
We assume familiarity with the \emph{text register machine} formalism $\one\hash$ in this section.
(However, one does not need to know much about that.  It would be possible to present a model
in the style of the Kleene PCA from Section~\ref{section-PCA}, and we do so in Section XXXX.)
The universe of the model is the set $W$ of words on $\set{\one,\hash}$.
We interpret $\eee$ by the empty word $\eps$,
$\ddd$ by the program {\sf diag} of $\one\hash$, 
 $+$ by concatenation of words, and 
the application operation $\app$ by 
\begin{equation}\label{problematic}
p \app x = \semantics{p}(x)
\end{equation}
That is, we take the word $p$, consider it as a program, and run that program on the word $x$.
We assume that $\eps$ is a program, and that it computes the identity function.
(We omit the precise formalization of $ \semantics{p}(x)$ for the time being.)
It is easy to check that the first three equations hold ``on the nose'':
$(W,+,\eps)$ is a monoid.  In addition, $\eee \app x = x$ and 
$\ddd \app\eee = \eee$ also hold.   The remaining equations are 
$(x + y ) \app z =  y \app (x \app z) $
and 
$(\ddd \app x) \app y   =  x \app (y + x)$.  These are problematic because the interpretation in (\ref{problematic})
is \emph{partial}.   We do have the required equations when we take into account the partiality:
\[
\semantics{x + y }(z)  \simeq  \semantics{y}(\semantics{x}(z))
\quad\mbox{ and } 
\quad
\semantics{\semantics{\mbox{\sf diag}}(x)}(y)
 \simeq  \semantics{x}(y + x)
\]
We write $\simeq$ in the usual sense of computability theory:
if one side is defined, then so is the other, and we have equality.
But again, in this model we have no a priori reason to think that the two sides above are defined.
In general, they are \emph{not defined}: since $\one\hash$
includes programs
for the empty partial function.  (In fact, it 
 is a Turing-complete formalism.)  
}

\subsection{Rewriting system and normal forms}
\label{section-pickup}

A \emph{rewrite rule} is a pair $(t,u)$ of terms, possibly allowing variables.
It differs from an equation in that it contains a left-to-write orientation.

We return to the set $E$ of equations in Figure~\ref{fig-1}.
We construct a rewriting system  $\RR$ by taking the pairs $(\ell,r)$
whenever $\ell = r$ is an equation in the system.
That is, we orient all equations in Figure~\ref{fig-1}  left-to-right, as they are written.
In particular, we orient the associative law for $+$ thus:  $(x + y) + z \to x + (y+z)$.

\begin{proposition}
\label{d-term-conf}
This term-rewriting system   $\RR$ is terminating and confluent.
\end{proposition}

To say that $\RR$ is terminating means there are no infinite sequences of applications of the rules.
To say that it is 
confluent:  If $t$ rewrites${}^*$ to both $u$ and $v$, 
then $u$ and $v$ can further rewrite${}^*$  to a common term $w$.

\takeout{
(VAR x y z)
(RULES
 add(e,y) -> y
 add(y,e) -> y
 add(add(x,y),z) -> add(x,add(y,z))
 app(add(x,y),z) -> app(y,app(x,z))
 app(app(d,x),y) -> app(x,add(y,x)) 
 app(e,x) -> x
 app(d,e) -> e
)
}

The termination and confluence results for term rewriting system in this paper are discussed briefly in Appendix A.

\takeout{
\begin{proof}
Both termination and confluence were checked with several programs 
run on the internet. 
(The details, at least some of them, are hidden in the source text.)
For the termination, we also can use the LPO  $\app > \ddd \sim + \sim \eee$.
The confluence is harder to talk about.
\end{proof}
}

It follows from Proposition~\ref{d-term-conf}
and Theorem~\ref{theorem-MG}
that every term $t$ has a unique \emph{normal form} $\nf(t)$.
By Theorem~\ref{theorem-MG}, the word problem for ground terms in this
signature is decidable, the initial algebra is the set of normal forms, and this 
algebra is computable.

\begin{defn} A term $t$ is a \emph{quine} if 
it is a normal form and 
$t\app \eee  \equiv t$.
Terms $t$ and $u$ are \emph{twins} if they are distinct normal forms
such that $t\app \eee \equiv u$ and $u\app \eee\equiv t$.
\end{defn}

\begin{example} [Standard presentation of quines]
The main example is $\ddd\app\ddd$. Note that
\[
(\ddd\app\ddd)\app \eee \to
\ddd\app(\eee + \ddd)  \to \ddd \app \ddd.
\]
In addition, $\eee$ is trivially a quine.
\label{ex-quine}
\end{example}

\begin{example}
The term $t = \ddd\app(\ddd + \ddd)$ is a normal form.
But $t\app \eee$ is not a normal form, and indeed we have
\[
t \app \eee \to (\ddd + \ddd)\app(\eee + (\ddd + \ddd)) \to 
\ddd \app (\ddd \app (\ddd + \ddd)).
\]
This last term is a normal form.

The term $u = (\ddd + \ddd)$ is a normal form.   But $u\app e$ is not a normal form,
and instead $u\app e \to \ddd\app(\ddd\app \eee) \to \ddd\app \eee \to  \eee$. 
\end{example}

The main results in this section are the characterization of normal forms,
Proposition~\ref{prop-characterization-nfs}, and the resulting results on quines and twins
in this language, Theorems~\ref{theorem-quines}
and~\ref{prop-no-cycles}.

\begin{proposition}\label{prop-nf-ddd}
If the term $t$ is in normal form,
then every subterm of $t$ of the form $u\app v$ has $u = \ddd$.
In particular, if $t$ is an $\app$-term which is a normal form,
then there is a normal form $u$ such that $t = \ddd\app u$.
\end{proposition}

\begin{proof}
 By induction on the term $t$.  We can assume that $t$ is a term in normal form,
 and that $t$ is an $\app$-term $u\app v$.
  If $u = \eee$, then $t$ would not be a normal form.   The same holds if $u$ were a $+$-term.
  So $u$ must either be $\ddd$ (and we are done), or $u$ is an $\app$-term.  Since $t$ is in normal form, so is $u$.
By induction hypothesis,
 $u$ is of the form $\ddd\app \www$.  But then $t$ is $(\ddd\app \www) \app v$, and this is a redex.
  So we have a contradiction in this case.
\end{proof}

Proposition~\ref{prop-nf-ddd} provides a necessary condition for a term to be a normal form:
in every application subterm $u\app v$, the term $u$ must be $\ddd$.   This condition is sufficient,
as the next result shows.   We first introduce some notation for later use.  

\begin{defn}
Let $N$, $N_{+}$, $N_{\app}$ be the smallest sets of terms such that
\begin{enumerate}
\item $\ddd,\eee\in N$.
\item $N_{+} \cup N_{\app} \subseteq N$.
\item If $t\in N\setminus\set{\eee}$, then $\ddd \app t$ belongs to $N_{\app}$.
\item If $n\geq 2$ and
$t_1, \ldots, t_n\in N_{\app}\cup\set{\ddd}$,
then $t_1 + (t_2 + \cdots + (t_{n-1} + t_n)\cdots ) \in N_{+}$.
\end{enumerate}
\end{defn}

Observe that $N$ is the disjoint union of 
the three sets  $\set{\eee,\ddd}$, $N_{\app}$, and $N_{+}$.

\begin{proposition}
\label{prop-characterization-nfs}
$N$ is the set of normal forms.
\end{proposition}

\begin{proof}
It is easy to check that all terms in $N$ are normal forms.
In the other direction, we check
by induction on the term $t$ that (a) if  $t$ is a normal form and an $\app$-term,
then $t\in N_{\app}$,
and (b) if $t$ is a
normal form and a $+$-term, then $t\in N_{+}$.
The first assertion is stated in Proposition~\ref{prop-nf-ddd}.
For the second, suppose that we have a normal form
 $t = u+ v$.  Neither $u$ nor $v$ can be $\eee$,
since $t$ is a normal form.   And $u$ cannot be a $+$-term, due to our orientation
of associativity.   So $u$ is either $\ddd$ or an $\app$-term.
As for $v$,  if $v = \ddd$, then $t = u + v \in N_{+}$.
If $v$ is an $\app$-term, then $v\in N_{\app}$ by induction hypothesis,
and we are done.
If $v$ is a $+$-term, then by induction hypothesis it is 
$w_1 + (w_2 + \cdots + (w_{m-1} + w_n)\cdots )$
where each $w_i\in  N_{\app}\cup\set{\ddd}$.  Since $t
=u + (w_1 + (w_2 + \cdots + (w_{m-1} + w_n)\cdots ))$,
 $t\in N_{+}$.
\end{proof}

\begin{proposition} If $t$ is a sum of terms $u_1, u_2, \ldots, u_k$
 in any parenthesization, and $t \equiv \eee$,
then for all $i$,   $u_i \equiv \eee$.
\label{prop-sum-term}
\end{proposition}

\begin{proof}
We show by induction on $n$ that if any finite sum
of terms $u_1 + u_2+ \cdots + u_k$
 (in any parenthesization) rewrites to $\eee$
in $n$ steps, then $u_i \equiv \eee$ for $1\leq i\leq k$.
For $n = 0$, the result is immediate.  Assume our result for $n$,
and suppose that   $u_1 + \cdots + u_k$
 rewrites to $\eee$
in $n+1$ steps.  If the first step of rewriting involves subterms of some $u_j$,
obtaining $u'_j$, 
consider the resulting term $u_1 + \cdots + u'_j + \cdots + u_k$.
By induction hypothesis, $u_i \equiv \eee$ for all $i\neq j$, and also $u'_j \equiv \eee$.
But since $u_j \equiv u'_j$, we also have $u_j \equiv \eee$.  
The other option in the induction step is when we use one of the first three rules in 
Figure~\ref{fig-1}: $x + \eee = x$, $\eee + x = x$, or associativity of $+$.
In all cases, the result follows easily from induction hypothesis.
\end{proof}

\takeout{
\begin{proof}
Consider the the term $t + u$, and assume that we have a reduction of this term to $\eee$.
The first step of this reduction must use one of the following three rules:
$x+ \eee\to x$, $\eee + x \to x$, or $(x + y) + z\to x + (y+z)$. 
In the first case, we have $u = \eee$ and $t \equiv \eee$.  So we are done.
The same holds in the second case.
In the third case, $t\in N_{\app}$ is $x + y$, and $u$ is $z$.
We may assume that $u\neq \eee$, or else we could be in our first case,
using the Church-Rosser property.
\end{proof}
}

\begin{proposition}\label{prop-not-a-plus}
For all normal forms $t\neq \eee$ and all normal forms $u$:
\begin{enumerate}
\item $\eee$ does not occur in $t$.
\label{part-no-e}
\item  $\nf(t\app u)\in \set{\eee}\cup N_{\app}$.
\label{not-a-plus-2}
\item If   $\nf(t\app u) = \eee$, then $u =\eee$.
\label{not-a-plus-3}
\end{enumerate}
\end{proposition}

\begin{proof} 
Part (\ref{part-no-e}) follows immediately from Proposition~\ref{prop-characterization-nfs}.
We show the other two parts (together) by induction on 
the term $t$.

For $t = \eee$, there is nothing to do.
For $t= \ddd$, $t + u$ cannot rewrite to $\eee$.
In the other parts, 
recall that if $u$ is a normal form and $u\neq \eee$, then
$\ddd \app u\in N_{\app}$.

Suppose that $t$ is an $\app$-term, say $t = \ddd\app v$.
Assume that $t$ is a normal form and hence is not $\eee$.
Assume  parts (\ref{not-a-plus-2}) and (\ref{not-a-plus-3})   for all subterms of $t$, and fix $u$.
Then $t\app  u = (\ddd\app v)\app u \equiv v\app (u+v)$.
Our induction hypothesis applies to $v$, since it is a 
subterm of $t$.
So 
\[ \nf(t\app u) = 
\nf(v\app (u+v))\in \set{\eee}\cup N_{\app}.
\]
And if $ \nf(t\app u) =  \eee$, then by induction hypothesis on $v$,
$u+v \equiv \eee$.
By Proposition~\ref{prop-sum-term},
 $u = \eee$.  This is what we want.

Finally, suppose that $t$ is a $+$-term normal form, say $t = v_1+ (v_2 + v_3)$.
(For $n \geq 3$ in the definition of $N_{+}$, the argument is similar.) Fix $u$.
Assume that $\nf(t\app u) =\eee$.
Then 
\[ t\app u =
( v_1+ (v_2 + v_3))\app u  \equiv  v_3\app( v_2\app( v_1\app u)).
\]
Our induction hypothesis applies to $v_1$, $v_2$, and $v_3$.
We have 
$\nf( v_3\app( v_2\app( v_1\app u)))=\eee\in \set{\eee}\cup N_{\app}$.
By applying our induction hypothesis to 
$v_3$, we see that $ v_2\app( v_1\app u) = \eee$.  Then we apply the induction 
hypothesis to $v_2$, and then $v_1$.  
We see that $u = \eee$, as desired. 
\end{proof}

\takeout{
\begin{proposition}\label{prop-not-a-plus}
For all normal forms $t$, $\nf(t\app \eee)\in \set{\eee}\cup N_{\app}$.
\end{proposition}

\begin{proof}
By induction on $t$.

If $t\in\set{\ddd,\eee}$,
then    $\nf(t\app \eee) = \eee$.

If $t$ is a $+$ term, say $x + y$ with $x$ and $y$ normal forms which are not $\eee$, 
then since $(x+y) \app \eee \to y\app( x\app \eee)$,
we see that $\nf(t\app \eee)$  is 
$\nf(y\app( x\app \eee))$.
If $x$ is $\ddd$, then this is $\nf(y\app \eee)$, and by induction hypothesis, this is not a $+$-term.
If $x$ is a $+$-term, then $t = x + y$ is not in normal form.
If $x$ is $\ddd\app z$ with $z$ a normal form,
then 
\[
\nf(t\app e) = \nf(y \app ((d\app z)\app \eee))) = \nf(y\app (z\app z)))
\]
This last term $y\app (z\app z)$ is a normal form, since $y$ and $z$ are.

If $t\in N_{\app}$, then write $t$ as $\ddd\app u$, where  $u$ is a normal form and is not
$\eee$.
Moreover, \[ \nf(t\app \eee) = \nf((\ddd \app u)\app \eee) = \nf(u \app (\eee + u) )
= \nf(u \app u) = 
u \app u.\]
Since $u\app u$ is a normal form and normal forms are unique, we have shown that 
$\nf(t\app \eee)$ is not a $+$-term.
\end{proof}
}

\subsection{Characterization of quines}

\begin{defn}
The \emph{$\ddd$-count} of a term $\dc(t)$ is 
the number of $\ddd$'s in it.
It is given by the following recursion:
\[ 
\begin{array}{lcl}
\dc(\eee) & = &0 \\
\dc(\ddd) & = &1 \\
\end{array}
\qquad
\begin{array}{lcl}
\dc(t+u) & = & \dc(t) +\dc(u)\\
\dc(t\app u) & = & \dc(t) +\dc(u)\\
\end{array}
\]
\end{defn}

\begin{proposition} 
\label{prop-only-dcts}
The only normal form $t$ with $\dc(t) = 0$ is $\eee$,
and the  only normal form $t$ with $\dc(t) = 1$ is $\ddd$.
\end{proposition}

\begin{lemma}
\label{lemma-dc}
Let $t$ be a term in which $\eee$ does not occur (not necessarily a normal form).
Then $\dc(\nf(t)) \geq \dc(t)$.
\end{lemma}

\begin{proof}
Fix $t$ with no occurrences of $\eee$.
We go from $t$ to its normal form by applying the rewrite rules.
  Each application of a rule maintains the $\dc$, with two
exceptions: $\ddd\app\eee \to \eee$, and 
$(\ddd \app x) \app y \to x \app (y + x)$. The first of these rules cannot apply
since we start with a term having no occurrences of $\eee$.
As for the rule $(\ddd \app x) \app y \to x \app (y + x)$, we lose 
the first $\ddd$ on the left, but this is made up by the extra $x$ on the right.
And $x$, like all subterms of $t$, must have at least one $\ddd$.  So
in going from $t$ to $\nf(t)$, we perform rewrites which maintain or increase 
the $\ddd$-count.
\end{proof}

\begin{remark}
The hypothesis that $\eee$ does not occur in $t$ is needed.
For example $\ddd\app \eee$ has a $\ddd$-count of $1$,
yet its normal form is $\eee$.
\end{remark}

\begin{theorem}\label{theorem-quines}
 [uniqueness of non-empty quines]
If $t$ is a
quine, 
then either $t=\eee $ or else $t = \ddd \app \ddd$.
\end{theorem}

\begin{proof}
Let $t$ be a normal form such that $\nf(t \app \eee ) = t$.
Let us also assume that $t$ is not $\eee$.
 Then $t$ cannot be $\ddd$, since $\ddd\app \eee = \eee$.
And $t$ cannot be a $+$-term by Proposition~\ref{prop-not-a-plus}.
So $t$ is an $\app$-term, say $\ddd\app u$ 
with $u$ a normal form and $u\neq \eee$
(see Proposition~\ref{prop-nf-ddd}).
Now
\[
\ddd\app u =
t =
\nf(t\app \eee) =\nf((\ddd\app u)\app e) = \nf(u\app u).
\]
Notice that $u\app u$ need not be a normal form.
But since $u\neq \eee$ is a normal form, $\eee$ does not occur in $u$
by Proposition~\ref{prop-not-a-plus}(\ref{part-no-e}).
Hence $\eee$ does not occur in $u\app u$.
Applying the $\ddd$-count function
and using  Lemma~\ref{lemma-dc}, we see that 
\[ 1 + \dc(u) = \dc(\ddd\app u) = \dc(\nf(u\app u)) \geq \dc(u\app u) = 2\cdot \dc(u).\]
Therefore $\dc(u) \leq 1$.  It follows that $u$ is either $\ddd$ or $\eee$, 
by Proposition~\ref{prop-only-dcts}.
As we know, $u\neq \eee$.
Thus $u = \ddd$, and so $t = \ddd\app u = \ddd\app \ddd$, as desired.
\end{proof}

\takeout{
\begin{proposition}\label{prop-no-twins} \marginpar{the next result is better, and this one is not needed.}
In the language of this section, there are no twins.
\end{proposition}

\begin{proof}
Assume towards a contradiction that $t$ and $u$ are twins.
It is clear that neither is $\eee$.  By 
Proposition~\ref{prop-not-a-plus}(\ref{not-a-plus-2}), neither is a $+$-term.
Thus, they are both $\app$-normal forms.
Suppose that 
$t = \ddd\app s$ and $u = \ddd\app v$.
Then $s\app s \equiv d \app v$, and $v \app v \equiv d \app s$.
As in Theorem~\ref{theorem-quines}, $1 + \dc(v) \geq 2 \dc(s)$, and $1 + \dc(s) \geq 2 \dc(v)$.
Since $\dc(s)$ and $\dc(v)$ are natural numbers, we see that either
$\dc(s) = \dc(v) = 0$, or else $\dc(s) = \dc(v) = 1$.
In the first case, $s$ and $v$ must be $\eee$, since this is the only normal form with $\ddd$-count $0$.
But then $t$ and $u$ are both $\eee$, and in particular, they are the same.
In the second case, $s$ and $v$ must be $\ddd$, and so $t$ and $u$ are both $\ddd\app\ddd$; again they are the same.
\end{proof}
}

\begin{proposition}\label{prop-no-cycles}
In the language of this section, there are no cycles of length $\geq 2$.
\end{proposition}

\begin{proof}
Let $n\geq 2$.  
In this proof,  the letter $i$ ranges over $\set{0,1,\ldots, n-1}$, and $i + 1$ taken $\textrm{mod } n$.
Assume that $t_0$, $\ldots$, $t_{n-1}$ are distinct normal forms, and $t_i \app \eee \equiv t_{i+1}$.
It is clear that no $t_i$ is $\eee$.  By 
Proposition~\ref{prop-not-a-plus}(\ref{not-a-plus-2}), no $t_i$ is a $+$-term.
Thus, each $t_i$ is an $\app$-normal form.  So we have terms 
$u_0$, $\ldots$, $u_{n-1}$ in normal form such that $t_i = \ddd \app u_i$.
Since $t_i\neq \eee$, we also have $u_i\neq \eee$.
For all $i$, 
\[
\ddd\app u_{i+1} = 
t_{i+1} \equiv t_i\app \eee =(\ddd\app u_i)\app \eee \equiv u_i \app u_i
\]
Thus $\ddd\app u_{i+1} = \nf(u_i \app u_i)$.  So 
\[
1 + \dc( u_{i+1} ) = 
\dc(\ddd\app u_{i+1}) = \dc( \nf(u_i \app u_i)) \geq
\dc(u_i\app u_i) = 2 \dc(u_i)
\]
Taking the sum of the last line for $i = 0,\ldots, n-1$ gives
\[
n + (\dc(u_0) + \cdots + \dc(u_{n-1})) \geq 2(\dc(u_0) + \cdots + \dc(u_{n-1}))
\]
Hence $\dc(u_0) + \cdots + \dc(u_{n-1})\leq n$.  Since each $\ddd$-count is a natural number, 
and none can be $0$, they are all $1$.  But then all $u_i$ are $\ddd$.  It follows that for all $i$,
$t_{i+1} = \ddd\app\ddd$.   This contradicts the assumption that the terms 
are $t_0$, $\ldots$, $t_{n-1}$  are distinct normal forms and  $n\geq 2$.
\end{proof}

\subsection{Further details on a computable model of the equations
which treats $\app$ as Kleene application}
\label{section-postpone}

In this section,
we continue the discussion of a model of our set $E$ equations that we started in Section~\ref{plainterms}. 
The goal is to have a model where $t\app u$ is interpreted as $\varphi_{t}(u)$ and where $+$ is concatenation of words or 
something close. (The reason that we say ``something close'' here is that the interpretation of $+$ in our model below
is not literally concatenation, due to the treatment of the comma.)
This stands in contrast with the term model of $E$, where the interpretation of the application function
on two terms $t$ and $u$ is simply the term $t\app u$.

We are discussing the ``programming language'' $P$ presented as pseudocode.
This is mainly for expository purposes.   The results here apply to the mathematically precise language $\one\hash$
presented by the author in~\cite{Moss06} and also implemented in courseware on computability theory.
The main things that one would need to know about $\one\hash$ are that (i) it is Turing-complete;
(ii) there are explicit programs corresponding to the constants mentioned in this paper;
(iii) programs are strings, and the concatenation of any two programs is a program (so that $+$ is associative);
(iv) the equational laws under study hold.

We write $T$ for the set of terms built from $\ddd$, $\eee$, $+$ and $\app$.
We are only interested in the ground terms, so there are no variables.
 We have a partial function $\semalt{\ }:T \to P$ given by
\[
\begin{array}{lcl}
\semalt{\ddd}  & = & \diag\\
 \semalt{\eee} & = & \eps\\
  \end{array}
  \qquad\qquad
 \begin{array}{lcl} 
 \semaltparens{t+u} & = & \semalt{t} + \semalt{u}\\
  \semaltparens{t\app u} & = &\semantics{\semalt{t}}(\semalt{u})\\
 \end{array}
 \]
Here  $\eps$ is the empty word, and on the right in the third equation, 
 we
 again  use $+$ for concatenation
 in $P$.
 The reason we emphasize that this is a partial function is that on the right we have application of a program to 
 an argument, and in general this is partial.   
 The main point is  that
 nevertheless,
  this function $\semalt{\ }$ is total.
  That is application of a program to an argument is in general partial, but for programs in the image of 
  the interpretation map ${}^*$, we need not worry.
Let 
\[\begin{array}{lcl}
A_0 & = & \set{x : x \mbox{ is a program of $P$}}\\
A_{n+1} & = & \set{x \in A_n : \mbox{for all $y\in A_n$, $\semantics{x}(y)$ is defined and belongs to $A_n$}}\\
\Aomega & = & \bigcap_n A_n
\end{array}
\]
Note that $A_1$ is the set of programs which compute total functions
(and indeed output programs), $A_2$ is the set of programs which
take elements of $A_1$ to elements of $A_1$, etc.
Observe that $A_0 \supseteq A_1 \supseteq A_2  \supseteq \cdots $.  In fact, the inclusions are all strict.

\begin{proposition}\label{prop-close-plus}
Each set $A_n$ is closed under $+$, where $+$ is concatenation of programs.
\end{proposition}

\begin{proof}
$A_0$ is closed under $+$ because the concatenation of two programs is always a program.
For $n+1$, let $x, y\in A_{n+1}$, and let $z\in A_n$.
Then $\semantics{x+y}(z) \simeq \semantics{y}(\semantics{x}(z))$.
Since $x\in A_{n+1}$, $\semantics{x}(z)$ is defined and belongs to $ A_n$.
And then since $y\in A_{n+1}$, $ \semantics{y}(\semantics{x}(z))$
 is defined and belongs to $A_n$.
 Thus $\semantics{x+y}(z)$ has this same property.
\end{proof}

 \begin{lemma}
 For all $t\in T$, $\semalt{t}$ is defined and is an element of $\Aomega$.
 In particular, $\semalt{\ }$ is a total function.
 \end{lemma}
 
 \begin{proof}
By induction on $T$.  First, $\semalt{\eee} = \eps$ belongs to $\Aomega$, 
since for $x\in A_n$, $\semantics{\eps}(x) = x\in A_n$.

Let us show next that $\semalt{\ddd} = \diag$ belongs to 
 $A_n$ for all $n$.
We use induction on $n$.
This is clear for $n =0$ since $\diag$ is a program.
Assume that $\diag\in A_n$.
To check that $\diag \in A_{n+1}$, let $x\in A_n$; we show that $\semantics{\diag}(x)\in A_n$.
If $n = 0$, this again is clear because $\semantics{\diag}$ is a total function on programs.
So we assume that $n >0$.
Let $y\in A_{n-1}$.
Now
\[
\semantics{\semantics{\diag}(x)}(y) = \semantics{x}(y+x)
\]
Since $x$ and $y$ belong to $A_{n-1}$, the program $y+x$ also belongs to $A_{n-1}$
by Proposition~\ref{prop-close-plus}.
Since $x\in A_n$, $\semantics{x}(y+x)\in A_{n-1}$.  Therefore, 
$\semantics{\semantics{\diag}(x)}(y) \in A_{n-1}$.  Since $y$ is arbitrary in $A_{n-1}$, 
$\semantics{\diag}(x) \in A_n$.  And since $x$ is arbitrary in $A_n$, $\diag\in A_{n+1}$.
This completes the induction and shows that $\diag\in \Aomega$.

Next, assume about $t$ and $u$ that $\semalt{t}$ and  $\semalt{u}$
belong to $\Aomega$.    Using Proposition~\ref{prop-close-plus}, we see that 
 $\semaltparens{t+u} = \semalt{t} +\semalt{u}$  belongs to $\Aomega$.
 
 Finally, assume about $t$ and $u$ that $\semalt{t}$ and  $\semalt{u}$
belong to $\Aomega$.  We show that $\semaltparens{t\app u}$ also belongs to $\Aomega$.
Fix $n$.  Then  $\semalt{u}\in A_n$ and  $\semalt{t}\in A_{n+1}$.
So  $\semalt{t}(\semalt{u})\in A_n$.  This for all $n$ shows what we want:
 $\semalt{t}(\semalt{u})= \semaltparens{t\app u} \in \Aomega$.
\end{proof}

\takeout{
\begin{theorem} For terms $t$ and $u$, $t\equiv u$ iff the functions 
 $\semantics{\semalt{t}}$ and  $\semantics{\semalt{u}}$ are the same.
\end{theorem}

\begin{proof}
We prove by induction on the complexity of $t$ and $u$ that if $t$ and $u$ are distinct 
normal forms, and if 
$x$, $x'$, $y$, and $y'$ are normal forms with
\[\begin{array}{llcl}
& \semantics{\semalt{t}}(\semalt{x})&=&  \semantics{\semalt{u}}(\semalt{y})\\
\mbox{and}  & \semantics{\semalt{t}}(\semalt{x'}) &=& \semantics{\semalt{u}}(\semalt{y'}) \\
\end{array}
\]
then $x = x'$ and $y = y'$.

Here is the induction step for terms $\ddd\app t$ and $\ddd\app u$.   We assume that 
\[\begin{array}{llcl}
& \semantics{\semalt{\ddd\app t}}(\semalt{x})&=&  \semantics{\semalt{\ddd \app u}}(\semalt{y})\\
\mbox{and}  & \semantics{\semalt{\ddd\app t}}(\semalt{x'}) &=& \semantics{\semalt{\ddd \app u}}(\semalt{y'}) \\
\end{array}
\]
That is, 
\[\begin{array}{llcl}
& \semantics{\semalt{t}}(\semalt{x+t})&=&  \semantics{\semalt{u}}(\semalt{y+u})\\
\mbox{and}  & \semantics{\semalt{t}}(\semalt{x'+t}) &=& \semantics{\semalt{u}}(\semalt{y'+u}) \\
\end{array}
\]
So by induction hypothesis, $x+t = x' + t$, and $y  + u = y'+u$.  It follows that $x = x'$ and $y = y'$.

Here is the induction step for terms $t_1 + t_2$ and $u_1 + u_2$.
 We assume that $t_1 + t_2 \neq u_1 + u_2$.  So either $t_2 = u_2$ and $t_1\neq u_1$,
 or else $t_2\neq 2_2$.   We have
\[\begin{array}{llcl}
& \semalt{t_1 + t_2}(\semalt{x})&=&  \semalt{u_1 + u_2}(\semalt{y})\\
\mbox{and}  & \semalt{t_1 + t_2}(\semalt{x'}) &=& \semalt{u_1 + u_2}(\semalt{y'}) \\
\end{array}
\]
That is, 
\[\begin{array}{llcl}
& \semantics{\semalt{t_2}}( \semantics{\semalt{t_1}}(\semalt{x}))&=&  
\semantics{\semalt{u_2}}(\semantics{\semalt{u_1}}(\semalt{y}))\\
\mbox{and}  & \semantics{\semalt{t_2}}(\semantics{\semalt{t_1}}(\semalt{x'})) &=& 
\semantics{\semalt{u_2}}(\semantics{\semalt{ u_1}}(\semalt{y'})) \\
\end{array}
\]
If $t_2 = u_2$ and $t_1\neq u_1$, then since $ \semantics{\semalt{t_2}}$ is one-to-one, we are easily done by the induction hypothesis.
If $t_2\neq u_2$, the the induction hypothesis tells us that
\[\begin{array}{llcl}
& \semantics{\semalt{t_1}}(\semalt{x})&=&  \semantics{\semalt{u_1}}(\semalt{y})\\
\mbox{and}  & \semantics{\semalt{t_1}}(\semalt{x'}) &=& \semantics{\semalt{u_1}}(\semalt{y'}) \\
\end{array}
\]
If $t_1 \neq u_1$, then the induction hypothesis on this pair finishes this case.
And if $t_1 =u_1$, then the fact that  $ \semantics{\semalt{t_1}}$ is one-to-one implies that $x = y$ and $x' = y'$.

All the same reasoning applies with a larger sum.

Here is the induction step for terms $\ddd\app t$  and  $u_1 + u_2$.
 We assume that 
\[\begin{array}{llcl}
& \semantics{\semalt{\ddd\app t}}(\semalt{x})&=& \semalt{u_1 + u_2}(\semalt{y})\\
\mbox{and}  & \semantics{\semalt{\ddd\app t}}(\semalt{x'}) &=& \semalt{u_1 + u_2}(\semalt{y'}) \\
\end{array}
\]
That is 
\[\begin{array}{llcl}
& \semantics{\semalt{t}}(\semalt{x+t})&=&  
\semantics{\semalt{u_2}}(\semantics{\semalt{u_1}}(\semalt{y}))\\
\mbox{and}  &\semantics{\semalt{t}}(\semalt{x'+t}) &=& 
\semantics{\semalt{u_2}}(\semantics{\semalt{u_1}}(\semalt{y'})) \\
\end{array}
\]
If $t \neq u_2$, then our induction hypothesis on that pair tells us that
$x+t = x'+ t$, so $x = x'$; also  $\semantics{\semalt{u_1}}(\semalt{y}) = \semantics{\semalt{u_1}}(\semalt{y'}) $.
Thus, $\semalt{y} = \semalt{y'}$.  So $y = y'$.

\end{proof}

\begin{proof}
We prove by induction on the complexity of $t$ and $u$ that if $t$ and $u$ are distinct 
normal forms, and if 
$x$ and $y$ are normal forms with
\[\begin{array}{llcl}
& \semantics{\semalt{t}}(\semalt{x})&=&  \semantics{\semalt{u}}(\semalt{x})\\
\mbox{and}  & \semantics{\semalt{t}}(\semalt{y}) &=& \semantics{\semalt{u}}(\semalt{y}) \\
\end{array}
\]
then $x = y$.

Here is the induction step for terms $\ddd\app t$ and $\ddd\app u$.   We assume that 
\[\begin{array}{llcl}
& \semantics{\semalt{\ddd\app t}}(\semalt{x})&=&  \semantics{\semalt{\ddd \app u}}(\semalt{x})\\
\mbox{and}  & \semantics{\semalt{\ddd\app t}}(\semalt{y}) &=& \semantics{\semalt{\ddd \app u}}(\semalt{y}) \\
\end{array}
\]
That is, 
\[\begin{array}{llcl}
& \semantics{\semalt{t}}(\semalt{x+t})&=&  \semantics{\semalt{u}}(\semalt{x+u})\\
\mbox{and}  & \semantics{\semalt{t}}(\semalt{y+t}) &=& \semantics{\semalt{u}}(\semalt{y+u}) \\
\end{array}
\]
So by induction hypothesis, $x+t = y + t$.  It follows that $x = y$.

Here is the induction step for terms $t_1 + t_2$ and $u_1 + u_2$.
 We assume that $t_1 + t_2 \neq u_1 + u_2$.  So either $t_2 = u_2$ and $t_1\neq u_1$,
 or else $t_2\neq 2_2$.   We have
\[\begin{array}{llcl}
& \semalt{t_1 + t_2}(\semalt{x})&=&  \semalt{u_1 + u_2}(\semalt{x})\\
\mbox{and}  & \semalt{t_1 + t_2}(\semalt{y}) &=& \semalt{u_1 + u_2}(\semalt{y}) \\
\end{array}
\]
That is, 
\[\begin{array}{llcl}
& \semantics{\semalt{t_2}}( \semantics{\semalt{t_1}}(\semalt{x}))&=&  
\semantics{\semalt{u_2}}(\semantics{\semalt{u_1}}(\semalt{x}))\\
\mbox{and}  & \semantics{\semalt{t_2}}(\semantics{\semalt{t_1}}(\semalt{y})) &=& 
\semantics{\semalt{u_2}}(\semantics{\semalt{ u_1}}(\semalt{y})) \\
\end{array}
\]
If $t_2 = u_2$ and $t_1\neq u_1$, then since $ \semantics{\semalt{t_2}}$ is one-to-one, we are easily done by the induction hypothesis.
If $t_2\neq u_2$, the the induction hypothesis tells us that
\[\begin{array}{llcl}
& \semantics{\semalt{t_1}}(\semalt{x})&=&  \semantics{\semalt{u_1}}(\semalt{x})\\
\mbox{and}  & \semantics{\semalt{t_1}}(\semalt{y}) &=& \semantics{\semalt{u_1}}(\semalt{y}) \\
\end{array}
\]
If $t_1 \neq u_1$, then the induction hypothesis on this pair finishes this case.
And if $t_1 =u_1$, then the fact that  $ \semantics{\semalt{t_1}}$ is one-to-one implies that $x = y$ and $x = y$.

All the same reasoning applies with a larger sum.

Here is the induction step for terms $\ddd\app t$  and  $u_1 + u_2$.
 We assume that 
\[\begin{array}{llcl}
& \semantics{\semalt{\ddd\app t}}(\semalt{x})&=& \semalt{u_1 + u_2}(\semalt{x})\\
\mbox{and}  & \semantics{\semalt{\ddd\app t}}(\semalt{y}) &=& \semalt{u_1 + u_2}(\semalt{y}) \\
\end{array}
\]
That is 
\[\begin{array}{llcl}
& \semantics{\semalt{t}}(\semalt{x+t})&=&  
\semantics{\semalt{u_2}}(\semantics{\semalt{u_1}}(\semalt{x}))\\
\mbox{and}  &\semantics{\semalt{t}}(\semalt{x+t}) &=& 
\semantics{\semalt{u_2}}(\semantics{\semalt{u_1}}(\semalt{y})) \\
\end{array}
\]
If $t \neq u_2$, then our induction hypothesis on that pair tells us that
$x+t = x+ t$, so $x = x$; also  $\semantics{\semalt{u_1}}(\semalt{x}) = \semantics{\semalt{u_1}}(\semalt{y}) $.
Thus, $\semalt{x} = \semalt{y}$.  So $y = y$.

\end{proof}
}

\begin{example} [Yang~\cite{yang}]
Let terms $p_n$ be defined as follows:
\[
\begin{array}{lcl} 
p_0 & = & \ddd \app (\ddd+ \ddd) \\
p_{n+1} & = & p_n \app \eee
\end{array}
\]
Yang observed that this gives a sequence of terms $p_0 \to p_1 \to \cdots $
which are different, where the arrow is from (\ref{express}).
Here is a bit more on this.
First, one shows that 
$\nf(p_n) \neq \ddd\app\ddd$ for all $n$.
Since $\to$ has no cycles, they must all be different.
But we are not interested in these as \emph{terms} but rather as actual programs.
That is, we fix an interpretation ${}^*\to \lang$ for some programming language $\lang$.
Each $p_n^*$ is a total function, as we know.  In particular, $p_n^*$ is well-defined.
To know that they are all different,
we would like to know that the  interpretation function  ${}^*\to \lang$ is one-to-one. 
But this is not something which we know at this time.  Instead, special purpose
counting arguments would be needed to show that the interpretations of the terms $p_n$
are all different programs.
\end{example}

\subsection{A (non-computable) model based on the natural numbers where again application is interpreted as Kleene application}

We exhibit a different model, one based on the natural numbers but where we identify numbers $m$ and $n$ if $\semantics{m} = \semantics{n}$
as partial functions.  This makes our model non-computable.

Fix numbers 
$i$,  $b$, $b'$, $\soneone$  and $d$
such that
\[
\begin{array}{lcl}
\semantics{i}(x) & = & x \\
\semantics{\semantics{\bfn}(x,y)}(z) & = &\semantics{x}(\semantics{y}(z)) \\
\semantics{\semantics{\bprimefn}(x,y)}(z) & = &\semantics{y}(\semantics{x}(z)) \\
\semantics{\semantics{\soneone}(x,y)}(z) & = & \semantics{x}(y,z)\\
\semantics{d}(x) & = & \semantics{b'}( \semantics{\soneone}(b,x),x)\\
\end{array}
\]
Then interpret $\eee$ by $i$, $\ddd$ by $d$,
$x + y$ by $\semantics{\bprimefn}(x,y)$, and
$\app$ by $n \app m = \semantics{n}(m)$.

The main reason that we want the identification that we mentioned above is that certain equations of interest
do not hold on the nose but only up to this identification.   For example, $\semantics{b}(x,y) \equiv \semantics{b'}(y,x)$, but the two sides
are not literally equal.

It is straightforward to verify the equations of interest.
The main verification: $(\ddd\app x)\app y = x \app(y + x)$.

\[
\begin{array}{lcl}
(\ddd\app x)\app y & = & 
\semantics{\semantics{d}(x)}(y) \\ 
& = &  \semantics{\semantics{b'}( \semantics{\soneone}(b,x),\emph{x})}(y) \\ 
& = & \semantics{\emph{x}}( \semantics{\soneone}(b,x)(y)) \\ 
& = &  \semantics{x}(\semantics{b}(x,y))\\ 
& \equiv &  \semantics{x}(\semantics{b'}(y,x))\\ 
& = &  x \app(y + x)\\
\end{array}
\]

\section{Adding a constant $\www$ for the ``write'' operation}
\label{section-equation-w}

We could at this point reconsider the constant symbol $\ddd$ and replace it with
the constant $\www$ along with the relevant equations.   If we did this, the
language would not be expressive enough for the kinds of questions which we ask
in this paper.  Specifically, there would not be any quine programs (see Corollary~\ref{noquines}).

We therefore \emph{add} $\www$ on top of the language which we saw in the previous section.
We adopt the equations in Figure~\ref{fig-2}.

\begin{remark}
Previously we had the equation $(\ddd \app x) \app y = x \app (y+x)$.
This equation is not found in  Figure~\ref{fig-2}.
However, it is derivable:
\[
\begin{array}{lcl}
(\ddd \app x) \app y 
= ((\www \app x)  + x) \app y
= x \app ((\www\app x) \app y)
= x \app (y + x).
\end{array}
\] 
\end{remark}

\begin{example} [Slattery's twins]
In contrast to what we saw in Proposition~\ref{prop-no-cycles} for the language with $\ddd$ but not $\www$, in the language 
that also has $\www$, there are twins.  
Let $p = \www\app (\ddd \app (\ddd+ \www))$ and $q = \ddd\app (\ddd + \www)$.
Then
\[\begin{array}{lclclclcl} 
p \app \eee &  = & (\www\app (\ddd \app (\ddd+ \www)))\app\eee& \equiv & \ddd\app (\ddd + \www) & = & q.\\
q\app \eee & = & (\ddd\app (\ddd + \www))\app\eee & \equiv &(\ddd + \www)\app (\ddd + \www)  & \equiv &
 \www\app (\ddd \app (\ddd+ \www)) & = & p.\\
\end{array}
\]
The normal forms  $t = \nf(p)$ and $u = \nf(q)$ are shown below.  They are distinct, so we have twins:
\[\begin{array}{lclclcl}
t  & = &
\www\app (\www\app \ddd) + \www\app(\www\app \www) + \www \app\ddd + \www\app \www.\\
u & = & 
\www\app\ddd  + \www\app \www 
+ \ddd + \www.\\
 \end{array}
\]
This pair of twins is due to Austin Slattery~\cite{Slattery}.
\end{example}

\begin{figure}[t]
\begin{mathframe}
\begin{array}{lcl}
x + \eee  & = & x \\
\eee + x & = & x \\
(x + y) + z & = &  x + (y+z) \\
(x + y ) \app z &= & y \app (x \app z) \\
\end{array}
\qquad\qquad
\begin{array}{lcl}
\ddd\app x & = & (\www \app x) + x \\
\eee\app x & = & x\\ 
\www \app (x +  y)& = & (\www \app x ) +  (\www \app y)\\
\www\app x\app y & = & y + x\\
\end{array}
\qquad\qquad
\begin{array}{lcl}
\www\app \eee & = & \eee\\
\\
\\
\\
\end{array}
\end{mathframe}
\caption{The set $E$ of equations used for the language with constants $\ddd$, $\eee$, and $\www$, and operations $+$ and $\app$.
\label{fig-2}}
\end{figure}

 \renewcommand{\arraystretch}{.3}

\begin{example}
Another pair of twins is
\[
\begin{array}{lcl}
t & = &  \www\app \www + \www\app (\www\app \ddd) +  \www + \www\app \ddd\\
u & = &  \www\app \www +  \www\app (\www\app \ddd)   +  \ddd \\
\end{array}
\]
Let us translate these to our stylized form of English commands $P$ from Section~\ref{plainterms}.
First, $t^*$ is
\[
\begin{array}{l}
\mbox{\emph{write ``write the instructions to write what is in the register''}},\\
\mbox{\emph{write ``write ``write the instructions to write what is in the register in front of it'' {}''}},\\
\mbox{\emph{write the instructions to write what is in the register}},\\
\mbox{\emph{write ``write the instructions to write what is in the register in front of it''}}
\end{array}
\]
Second, $u^*$ is
\[
\begin{array}{l}
\mbox{\emph{write ``write the instructions to write what is in the register''}},\\
\mbox{\emph{write ``write ``write the instructions to write what is in the register in front of it'' {}''}},\\
\mbox{\emph{write the instructions to write what is in the register in front of it}}
\end{array}
\]
\end{example}

 \renewcommand{\arraystretch}{1.5}
 
\begin{example} [Slattery's multiples]
Slattery also observed that his twins generalize to cycles of all lengths.
Here is the construction of a $3$-cycle.
Let 
\[
p = (\ddd + \www + \www)\app  (\ddd + \www + \www),
\]
and observe that  $p \equiv \www\app(\www\app(\ddd\app   (\ddd + \www + \www)))$.  Then
\[\begin{array}{lcl}
p \app \eee &\equiv &\www\app(\ddd\app (\ddd + \www + \www)) \\
(\www\app(\ddd\app (\ddd + \www + \www))) \app \eee & \equiv & \ddd\app (\ddd + \www + \www )\\
(\ddd\app (\ddd + \www + \www))\app \eee & \equiv & p
\end{array}
\]
\end{example}

\takeout{
\[\begin{array}{lcl}
p \app \eee &\equiv & \www \app (\www \app p) \\
( \www \app (\www \app p) ) \app \eee & \equiv & \www\app p \\
(\www \app p) \app \eee & \equiv & p
\end{array}
\]
}

\begin{defn} As we did with our previous set of equations, 
we construct a rewriting system using the relation $\ell \to r$ 
whenever $\ell = r$ is an equation in Figure~\ref{fig-2}.
We overload our notation to call this rewriting system $\RR$.
\end{defn}

\begin{proposition}
This term-rewriting system $\RR$ is terminating and confluent.
\end{proposition}

\takeout{
\begin{proof}
For the termination, use the LPO  $\ddd\sim\app > \www \sim + \sim \eee$.
\end{proof}
}

In a normal form, each application term $t\app u$ has $t = \www$.
Thus, every normal form $t$ is a finite sum of terms, each 
of the following three forms:
$\ddd$, $\www$, or $\www\app t'$, where $t'$ is again a normal form other than $\eee$.

\begin{theorem}
\label{theorem-characterize-Slattery}
Let $t$ and $u$ be 
normal forms
such that $t\app \eee \equiv u$ and $u\app \eee \equiv t$.
Then one of the following holds:
\begin{enumerate}
\item $t = \eee = u$.
\item 
$t = (\www\app\ddd) + \ddd = u$.  In other words,
$t \equiv \ddd \app \ddd  \equiv u$.
 \item
$t =  \www\app \www + \www\app (\www\app \ddd)+  \www + \www\app \ddd$,
and $u =  \www\app \www +  \www\app (\www\app \ddd)   +  \ddd $, 
or vice-versa.
\item 
$ t= \www\app(\www\app \ddd) + \www\app(\www\app\www) + 
 (\www\app \ddd) + (\www\app\www)$
 and 
 $ u = (\www\app \ddd) + (\www\app\www)+ \ddd + \www$,
 or vice-versa.  In other words, $t$ and $u$ are Slattery's twins.
\end{enumerate}
\end{theorem}

The proof is a long case-by-case analysis, and we lack the space for it here.

\takeout{
\begin{proof}
We assume throughout that neither $t$ nor $u$ is $\eee$,
for if one of these is $\eee$ then so is the other.

Let $n_t$ be the number of summands of $t$ which are $\ddd$ or $\www$,
let $m_t$ be the number of other summands of $t$.
Let $n_u$  and $m_u$ be defined similarly from $u$.

We also cannot have $m_t = 0$.
For if $m_t = 0$, then $t\app\eee \equiv \eee$.
For the same reason,  we cannot have $m_u = 0$.

We cannot have both $n_t = 0$ and $n_u = 0$.
For suppose that we did.
Then we can write 
\[ 
\begin{array}{lcl}
t &  = & (\www\app x_1) + \cdots + (\www\app x_k) \\
u & = & (\www\app y_1) + \cdots + (\www\app y_\ell)
\end{array}
\]
Applying each to $\eee$ shows that 
\[ 
\begin{array}{lcl}
t &  = &y_1 + \cdots + y_\ell \\
u &  = &x_1 + \cdots + x_k \\
\end{array}
\]
These are normal forms, in view of the omitted
parenthesization.
But then the height of the first summand of $t$ (namely $\www\app x_1$) is larger
than the height of the first summand of $u$ (namely $x_1$); and the same argument
shows the reverse.   So we have a contradiction.

Next, we show that we cannot have both $n_t \geq 2$ and $n_u \geq 2$.
This is actually similar to the last paragraph.

Much the same reasoning shows that we cannot have $n_t \geq 3$
(and similarly, we cannot have $n_u \geq 3$).

Let us show that\footnote{It might be better to 
make a claim here showing that if the first symbol of $t$  is a constant,
then the same holds for $u$, and also $n_t = n_u$.  This fact is what would be 
used in a few places, including right here.} 
the first term of $t$ cannot be $\ddd$ or $\www$.  
Assume towards a contradiction that it were.  In this argument, we are going 
to use $c$ as a variable ranging over $\set{\www,\ddd}$, and we call such terms 
\emph{constants}.
The only way to have the first term of $t$ a constant is if the first term of $u$
were a constant (and this will then be the only constant in $u$, since $n_u \leq 1$).
Indeed, if $t = c + x_1 + \cdots + x_n$, then $u = c' + (\www\app c) +
(\www\app x_1) + \cdots + (\www\app x_n)$.  Thus, the number of summands in
$u$ is $1$ more than the number of summands in $t$.
But then the same argument applies in reverse, and \emph{all} of the constants in
$t$ must appear at the front.
So $t$ is $t = c_1 + c_2 + c_3 + x_1 + \cdots + x_n$, 
with all $x$'s non-constants.  Moreover, the number of summands in $t$
is $3$ more than the number of summands in $u$.
So by looking at those numbers of summands, we have a contradiction. 

So the first term of $t$ must be $\www\app x$ for
some $x$. But then if $n_t\geq 3$, then the first term of $u \equiv t \app \eee$ is
$\www\app(\www\app x_1)$.  But this is $\www\app y_1$.  And then the first term of $t$
is $y_1$, since $\ldots$.

So at this point, we have a few options.  Omitting cases that are equivalent by symmetry:
\begin{enumerate}
\item $n_t = 0$ and $n_u = 1$.
Let us say that 
\[
\begin{array}{lcl} t &  = & (\www \app x_1) + \cdots (\www \app x_k)\\
u & = &  (\www\app y_1) + \cdots + (\www\app y_\ell)+ \www\\
\end{array}
\]
Then $t\app \eee = x_1 + \cdots + x_k$, and since this is a normal form,
$x_1 = \www\app y_1$.  But 
\[ t = u\app \eee =  (\www\app y_1) + \cdots + (\www\app y_\ell).\]
Since this last expression is a normal form, too, $\www\app y_1 = \www \app x_1$.
This means that $x_1 = y_1$.  So we have a contradiction.
 
\item $n_t = 1$ and $n_u = 1$.
By applying to $\eee$, it is easy to see that neither 
$t$ nor $u$ can be of the form $\www + (\www\app x)$ or 
$\ddd + (\www\app x)$.

Write $t = (\www\app x) + x'$ and $u = (\www\app y) + y'$,
where $x'$ and $y'$ are either $\www$ or $\ddd$.
In this case, neither can be $\www$ lest the other term not have a ``constant term''.
So  $t = (\www\app x) + \ddd$ and $u = (\www\app y) + \ddd$.
Applying these to $\eee$ gives $u = (\www\app x) + x$, and $t = (\www\app y) + y$.
Since the terms on the right are normal forms, we have $x = \ddd = y$, and so 
$t = (\www\app \ddd) + \ddd = u$.

\item $n_t = 0$ and $n_u = 2$.  Let
\[
\begin{array}{lcl}
t &  = &  (\www\app x_1) + \cdots +  (\www\app x_k)\\
u & = &  (\www\app y_1) + \cdots +  (\www\app y_\ell) + z' + z'\\
\end{array}
\]
where $z'$ and $z''$ are $\www$ or $\ddd$, and these terms can be mixed
with the previous ones. We first rule out the case that the first term of 
$u$ is $\www$ or $\ddd$.  This is done as in the paragraph above
which begins ``Much the same reasoning.'' In fact, the first part of 
the argument there applies here.

The number $k$ must be $\ell + 2$, and by easy counting, we see that 
$\ddd$ must be the third term in $u$, and $\www$ the last term.
In more detail, if $\ddd$ were the fourth or greater term, 
then $u\app\eee$ would add at least six terms, and only two ($\ddd$ and $\www$)
would disappear into $t$.   Similarly, if $\ddd$ were the first or second term in $u$,
then $u\app\eee$ would not gain two terms.   This shows that $\ddd$ is exactly the 
third term.  

We next claim  $\www$ must be the last
term in $u$. If not, let 
$j$ be such that $\www$ is the $j$th term in $u$.
The $(j+1)$st term in $u$ is  $\www\app  z_1$ for some $z_1$.
Let the $(j+1)$st term in $t$ be $\www\app  z_2$.
Then when we apply $t$ and $u$ to $\eee$, we see that
$z_1 = \www\app  z_2$, and $z_2 = \www\app  z_1$.
So $z_1 = \www\app  (\www\app  z_1)$. This is a contradiction.

So the form of $t$ and $u$ is
\[
\begin{array}{lclclclclclclclcl}
t & = & (\www\app x_1) &+& & & &  & & &  \cdots &+& (\www\app x_{k-1})
&+& (\www\app x_{k})    \\
u  & = & (\www\app y_1) &+&  (\www\app y_2) &+& \ddd & + & (\www\app y_3) 
& + & \cdots &+& (\www\app y_{k-2}) &+ &\www \\
\end{array}
\]
Again, $\ddd$ is the third term in $u$, so $\www\app y_1$ and
 $\www\app y_2$ are definitely present in $u$.
We next claim that the terms $ (\www\app y_3)$, $\ldots$, 
$ (\www\app y_{k-2}) $ must be missing from the sum above,
 so that $k = 4$.
If not, consider $\www\app x_{k-1}$ and $\www\app y_{k-2}$.
By applying $t$ and $u$ to $\eee$, we have $x_{k-1} = y_{k-2}$, and 
$\www\app y_{k-2} = \www\app x_{k-1}$.  Thus, 
$ y_{k-2} = \www\app y_{k-2}$, and this is a contradiction.

At this point, we see that $t$ and $u$ are of the following forms:
\[
\begin{array}{lclclclclcl}
t & = & (\www\app x_1) &+&  (\www\app x_2) &+& (\www\app x_3) &+&  (\www\app x_4)\\
u  & = & (\www\app y_1) &+&  (\www\app y_2) &+& \ddd & + & \www \\
\end{array}
\]
Apply $u$ to $\eee$  to see that 
\[
\begin{array}{lclclclclcl}
t & = & (\www\app (\www\app y_1)) & + &  (\www\app (\www\app y_2))
& + &  (\www\app y_1) &+&  (\www\app y_2)\\
u & = & x_1 & + & x_2  & + & x_3 & + & x_4\\
 \end{array}
\] 
From the last terms of $u$ just above,
 we see that $x_3 = \ddd$ and $x_4 = \www$.
From the last terms of $t$, we see that $y_1 = x_3$, and $y_2 = x_4$.
So we have 
\[
\begin{array}{lclclclclcl}
t & = &  (\www\app (\www\app \ddd)) & + &  (\www\app (\www\app \www)) &+& (\www\app \ddd) &+&  (\www\app \www)\\
u  & = & (\www\app \ddd) &+&  (\www\app \www) &+& \ddd & + & \www \\
\end{array}
\]
Thus, $t$ and $u$ are Slattery's twins.

\item $n_t = 1$ and $n_u = 2$.
We first claim that in both $t$ and $u$, the first term must be of the form $\www\app x$.
That is, the first term in both cannot be a constant.  The same argument
as in the paragraph ``Much the same reasoning'' applies; the only thing
that was used there was that $n_t \neq n_2$.

Let
\[ 
\begin{array}{lcl} t &   = &  (\www\app x_1) + \cdots +  (\www\app x_k) + x'\\
u &  = &  (\www\app y_1) + \cdots +  (\www\app y_\ell) + y' + y''\\
\end{array}
\]
where $x'$, $y'$, and $y''$ are either $\www$ or $\ddd$.  
(We have put these ``constant terms'' at the end, but they might be mixed in with 
the other summands.)
By evaluating $t\app\eee$ and comparing the first terms of the normal forms, we see that 
$\www\app x_1 = \www\app y_1$; thus $x_1 = y_1$.
But evaluating $u\app e$ then gives the first term of $t$ to be $\www\app(\www\app x_1)$.
This is a contradiction.

\end{enumerate}

This completes the proof of Proposition~\ref{prop-characterize-Slattery}.
\end{proof}
}

\begin{corollary}\label{noquines}
 In the language with only $\www$, $\eee$, $+$ and $\app$, there 
are no non-trivial quine programs.
\end{corollary}

\section{A next step: adding  constants $\uuu$ for the universal function
and $\sss$ for $s^1_1$}
\label{section-large}

Figure~\ref{fig-1} formulates a larger language than what we studied in the main body
of the paper.  It includes a constant for the $\sss^1_1$ function (expressed in curried form, 
as one would expect here) and two constants for different presentations of
the universal function in computability theory.  The most natural way to formulate this 
is via what we wrote as $\uuu$, with the equation $\uuu\app x \app y = x \app y$.
And for this paper, we also want a constant $\uuu_0$ representing running a program
on ``empty registers''.  The equation here is $\uuu_0 \app x = x \app \eee$.  

What we know about the associated rewriting system is that it is confluent,
and without the equation for $\uuu_0$, it is also terminating.  This means that 
the word problem is decidable for it. 
But adding $\uuu_0$ and its equation results in a term rewriting system which is
(surprisingly) non-terminating:
\[
\begin{array}{llllllllll}
   (\ddd \app (\ddd + \uuu))\app\eee
\to &  (\ddd + \uuu)\app (\eee + (\ddd + \uuu))
& \to \uuu \app (\ddd \app (\eee + (\ddd + \uuu)))
\\
&   & 
\to   \uuu \app (\ddd \app (\ddd + \uuu))
\to   (\ddd \app (\ddd + \uuu))\app\eee
\end{array}
\]
We do not know if the word problem remains decidable.

\takeout{
https://aprove.informatik.rwth-aachen.de/interface/submission

s = APP(u, app(d, add(e, add(x, y)))) evaluates to t =APP(y, app(x, add(e, add(x, y))))

APP(u, app(d, add(e, add(d, u)))) → APP(u, app(d, add(d, u)))
with rule add(e, y') → y' at position [1,1] and matcher [y' / add(d, u)]

APP(u, app(d, add(d, u))) → APP(app(d, add(d, u)), e)
with rule APP(u, x') → APP(x', e) at position [] and matcher [x' / app(d, add(d, u))]

APP(app(d, add(d, u)), e) → APP(add(d, u), add(e, add(d, u)))
with rule APP(app(d, x'), y') → APP(x', add(y', x')) at position [] and matcher [x' / add(d, u), y' / e]

APP(add(d, u), add(e, add(d, u))) → APP(u, app(d, add(e, add(d, u))))
with rule APP(add(x, y), z) → APP(y, app(x, z))

 add(e,y) -> y
 add(y,e) -> y
 add(add(x,y),z) -> add(x,add(y,z))
 app(add(x,y),z) -> app(y,app(x,z))
 app(app(d,x),y) -> app(x,add(y,x)) 
 app(e,x) -> x
 app(d,e) -> e
 app(x,e) -> app(u,x)
 
 termination shown at http://tfmserver.dsic.upv.es:8080/TRS.aspx
 
 Here is where to find the tools: https://www.jaist.ac.jp/~hirokawa/tool/?

 Confluence checked by 
 
 http://tfmserver.dsic.upv.es:8080/TRS.aspx
 
}

\section{Conclusion and Future Work}

We proposed several sets of equations related to ``self-expression'' by programs, or more generally to 
the expression of one program by another.
The overall goal was to have an equational axiomatization of very simple but very useful parts of 
computability theory, and then to study  the systems of equations which emerge.   The main
results concerned the equations in Figures~\ref{fig-1} and~\ref{fig-2}.  We observed that in both 
of these, the word problem is decidable, calling on results from term rewriting theory.
Indeed, the natural rewriting systems coming from these sets of equations are terminating and confluent.
Then we studied the normal forms in both systems, obtaining a few results on self-writing programs
and cycles.

The decidability results in this paper contrast with those in stronger systems like combinatory logic.
From our point of view, this is because CL usually starts with the $\sss$ and $\kkk$ combinators,
and thus it starts with everything one would need to represent every computable partial function.
What we did in this paper goes the other way, starting only with what is necessary to study the 
phenomena of interest.   What we lose in expressive power is (we hope) recovered in applicability
and interest of the results.  Be that as it may, what we are doing fits a pattern in logic and theoretical
computer science of studying ``light'' systems, where again one aims for logical systems which are 
both useable and tractable.

We were able to classify the quines and cycles in our equational logics.   The uniqueness of quine programs
is at first surprising.  Some have felt that it must be wrong, since in general there are no uniqueness 
results for anything in computability theory.   All of the constructions can be modified in small ways.
We are not saying that quine programs are unique, of course.  We are saying that in our equational logics
of either  Figures~\ref{fig-1} or~\ref{fig-2}, there is just one term $q$ such that $q\app \eee \equiv q$.
Moving to a real programming language $\lang$, or to indices of computable functions, means defining an 
interpretation ${}^*: T \to \lang$, where $T$ is the set of terms under discussion.  It is this interpretation function 
which is so non-unique.  There are many ways to modify one such function.  In a sense, what we are saying
about quine programs is that if one wants to write one using a \emph{fixed} diagonal function and allows
composition of programs, and nothing else, then there is only one way to do it.   Even if one adds the 
 ``write'' construct, there is just one quine program, up to equivalence in a simple equational logic.
 Moreover, adding that  ``write'' construct gives cycles of all lengths; these would not exist without ``write.''
 All of these results were based on term rewriting theory, and we know of no other way to obtain them.
 
Our results suggest a number of open questions.  

Returning to our discussion of quine programs in everyday programming languages,
we wonder whether 
a language which does \emph{not} allow programs to apply to other programs can possibly
have an interpretation of the diagonal operator $\diag$ in our sense.  But such a language 
would still have quine programs (and more).   So one wonders whether there is a different form 
of equational reasoning that could illuminate the self-referential aspects of such languages.
Admittedly, this question  is vague.

 We formulated a set of equations in Figure~\ref{fig-1}
and we know about the termination and confluence properties of
the associated term rewriting system.   
 (It is confluent, but the one equation $\uuu_0 \app x = x\app \eee$ messes up the termination.
 Without this, the system is terminating.) But we did not go further with this set of equations,
 and it is open to see whether characterization results like 
 Theorems~\ref{theorem-quines}
and~\ref{prop-no-cycles} hold for this bigger logic.


Another open question concerns the computable model which we discussed in Section~\ref{section-postpone}.
We showed that one can interpret terms in our first equational logic in a pseudocode language, and that 
the interpretation of every term is a total computable function (and more).  We conjecture that this same 
results holds for arbitrary programming languages; as soon as one has a Turing-complete language,
we believe that it should interpret our logical systems.  Moreover, the interpretation function which we studied
should be one-to-one.
One precise formulation would be that the partial combinatory algebra built from Kleene application should have
the initial algebra of our equations as a subalgebra, and Kleene application in  this subalgebra should be total.
We leave this for future work.

A final open question concerns extensional models of our equations.  In CL, extensionality is the principle that if $x$ is a variable, 
from $tx = ux$, one infers $t = u$.  This can be formulated for our systems.  We do not know if the decidability results
here extend to the system that also includes extensionality.

\section{Acknowledgments}
I thank the anonymous referees for criticisms of the presentation and of specific points in the paper.

I also am grateful to my computability students in the fall of 2022 for their enthusiasm for the study of the 
program expression graph, and especially to Austin Slattery and Yafei Yang for their work
and for their permission to mention it here.   I thank Peter Gerdes for discussion of proofs of the Recursion Theorem
which are not equational in the sense of this paper.  I am grateful to Catharine Wyss for her interest in this 
project and for many discussions of it.

I thank Neil Jones for a series of email discussions in the period  November 2022-January 2023.  
Neil passed away on March 26, 2023.  He was interested in this project as it pertains to some of his work
and to the Futamura projections coming from partial evaluation.   He himself was formulating ``small'' 
but expressive computational systems. He generously discussed matters with me in his last months.  He wrote,
``The bottom line seems to be that
     [my current work] emphasises interpreters and compilers in a many-type, many-language
        context, and the $\ldots$, universal program; while
[your draft] seems to emphasise self-generating programs and Kleene's 2nd Recursion
        and S-m-n Theorems.
I think there is a lot in common, even though the end goals seem rather different.''
I dedicate this paper to his memory.

\takeout{
\section{Computable Models}

\textbf{This section is probably trash.}

We present two (related) models for the calculus.
The first uses a fixed enumeration of the computable 
unary and binary partial
functions 
on the natural numbers as $\semantics{e}$, as $e$ ranges over
the natural numbers.
(The traditional notation is $\phi_e$.)
We fix such an enumeration, and we also fix numbers 
$i$, $b$, $b'$ and $\soneone$
such that
\[
\begin{array}{lcl}
\semantics{i}(x) & = & x \\
\semantics{\semantics{\bfn}(x,y)}(z) & = &\semantics{x}(\semantics{y}(z)) \\
\semantics{\semantics{\bprimefn}(x,y)}(z) & = &\semantics{y}(\semantics{x}(z)) \\
\semantics{w^*}(x) & = & \semantics{\soneone}(b,x)\\
\end{array}
\]
In other words
\[
\begin{array}{lcl}
\semantics{i} & = & \id \\
\semantics{\semantics{\bfn}(x,y)} & = &\semantics{x} \o \semantics{y} \\
\semantics{\semantics{\bprimefn}(x,y)} & = &\semantics{y} \o \semantics{x} \\
\end{array}
\]
To each term $t$ we associate a number $t^*$ by the following recursion:
\[
\begin{array}{lcl}
\eee^* & = & i \\
\www^* & = & w^* \\
(t \app u)^* & = & \semantics{t^*}(u^*)\\
(t + u)^* &  = & \semantics{\bprimefn}(t^*,u^*)\\
\end{array}
\]

\paragraph{The model} Define $\equiv$ on $N$ by 
\[
x \equiv y \quadiff \semantics{x} = \semantics{y}
\]
The domain of $\Model$ is the quotient $N/\!\!\equiv$.
The this relation is a congruence for the operations above, and so this quotient
interprets the signature.  We check that the equations are satisfied.
Let us introduce a little notation to help.  We write $\pair{t}$ for $\semantics{t^*}$.
Here is what we need to verify:
\[
\begin{array}{lcl}
\pair{x + \eee}  & = & \pair{x} \\
\pair{\eee + x} & = & \pair{x} \\
\pair{(x + y) + z} & = &  \pair{x + (y+z)} \\
\\
\end{array}
\qquad\qquad
\begin{array}{lcl}
\pair{(x + y ) \app z} &= & \pair{y \app (x \app z)} \\
\pair{\www \app (x +  y)}& = & \pair{(\www \app x ) +  (\www \app y)}\\
\pair{(\www\app x)\app y} & = & \pair{y + x}\\
\pair{\www\app \eee} & = & \pair{\eee}\\
\end{array}
\]
Since $\pair{x + y } = \pair{y}\o \pair{x}$ and $\pair{\eee}=\id$,
 the first three equations are trivial.
\[
((x + y)\app z)^* = \semantics{(x+y)^*}(z^*) = (\semantics{y'}\o \semantics{x^*})(z^*)
= \semantics{y^*}(\semantics{x^*}(z^*)) = (y@(x@z))^*
\]

In the remaining verifications, notice that 
$(\www\app x)^* = \semantics{\soneone}(b,x^*)$.

\[
\begin{array}{lcl}
(\www \app (x +  y))^* & = &  \semantics{\soneone}(\bfn,\semantics{\bprimefn}(x^*,y^*)) \\
\\
& = &
\semantics{\bprimefn}( \semantics{\soneone}(b,x^*),  \semantics{\soneone}(b,y^*))
\\
& = & 
\semantics{\bprimefn}((\www\app x)^*,(\www\app y)^*)\\
& = &
((\www \app x ) +  (\www \app y))^*
\end{array}
\]

\[
\begin{array}{lcl}
\pair{w\app (x+y)}(z) & = & ( \semantics{\soneone}(\bfn,\semantics{\bprimefn}(x^*,y^*)))(z) \\
& = & \semantics{\bfn}(\semantics{\bprimefn}(x^*,y^*),z)\\
& = & \semantics{\bfn}(\semantics{\bfn}(y^*,x^*),z)\\
& \equiv & \semantics{\bfn}(y^*,\semantics{\bfn}(x^*,z))\\
& = & 
\pair{\www\app y}(\pair{w\app x}(z)) 
\\
& = & \pair{(\www \app x ) +  (\www \app y)}(z) \\
\end{array}
\]

\[
\begin{array}{lcl}
((\www \app x)\app y)^*  & = & \semantics{(\www\app x)^*}(y^*)\\
& = & \semantics{\semantics{\www}(x^*)}(y^*)\\
& = & \semantics{\semantics{\soneone}(\bfn,x^*)}(y^*)\\
& = & \semantics{\bfn}(x^*,y^*) \\
& \equiv & \semantics{\bprimefn}(y^*,x^*) \\
& = & (y+x)^*
\end{array}
\]

\[
\begin{array}{lcl}
(\www\app \eee)^*(x) & = & \semantics{\semantics{\soneone}(\bfn, \eee^*)}(x) \\
& = & \semantics{\bfn}(i^*,x) \\
& \equiv & x
\\
& = & \semantics{\eee^*}(x)
\end{array}
\]

}

    \bibliographystyle{plainurl}
    \bibliography{refs}

\begin{thebibliography}{10}

\bibitem{bimbo}
Katalin Bimb\'{o}.
\newblock {\em Combinatory logic}.
\newblock CRC Press, Boca Raton, FL, 2012, ISBN: 9781439800003

\bibitem{BonfanteKM07}
Guillaume Bonfante, Matthieu Kaczmarek, and Jean{-}Yves Marion.
\newblock A classification of viruses through recursion theorems.
\newblock In S.~Barry Cooper, Benedikt L{\"{o}}we, and Andrea Sorbi, editors,
  {\em Computation and Logic in the Real World, Third Conference on
  Computability in Europe, CiE 2007, Siena, Italy, June 18-23, 2007,
  Proceedings}, volume 4497 of {\em Lecture Notes in Computer Science}, pages
  73--82. Springer, 2007.
\newline \href {https://doi.org/10.1007/978-3-540-73001-9_8}
  {\path{doi:10.1007/978-3-540-73001-9_8}}.

\bibitem{DiPaola}
Robert~A. Di~Paola and Alex Heller.
\newblock Dominical categories: recursion theory without elements.
\newblock {\em J. Symbolic Logic}, 52(3):594--635, 1987.
\newline \href {https://doi.org/10.2307/2274352} {\path{doi:10.2307/2274352}}.

\bibitem{AProVE}
J{\"{u}}rgen Giesl, Cornelius Aschermann, Marc Brockschmidt, Fabian Emmes,
  Florian Frohn, Carsten Fuhs, Jera Hensel, Carsten Otto, Martin Pl{\"{u}}cker,
  Peter Schneider{-}Kamp, Thomas Str{\"{o}}der, Stephanie Swiderski, and
  Ren{\'{e}} Thiemann.
\newblock Analyzing program termination and complexity automatically with
  {AProVE}.
\newblock {\em J. Autom. Reason.}, 58(1):3--31, 2017.
\newline \href {https://doi.org/10.1007/s10817-016-9388-y}
  {\path{doi:10.1007/s10817-016-9388-y}}.

\bibitem{Jones13}
Neil~D. Jones.
\newblock A {S}wiss pocket knife for computability.
\newblock In {\em Semantics, Abstract Interpretation, and Reasoning About
  Programs: Essays Dedicated to {D}avid {A}. {S}chmidt on the Occasion of his
  Sixtieth Birthday}, volume 129 of {\em Electron. Proc. Theor. Comput. Sci.
  (EPTCS)}, pages 1--17. EPTCS, 2013.
\newline \href {https://doi.org/10.4204/EPTCS.129.1}
  {\path{doi:10.4204/EPTCS.129.1}}.

\bibitem{Kleene38}
Stephen~C. Kleene.
\newblock On notation for ordinal numbers.
\newblock {\em Journal of Symbolic Logic}, 3(4):150--155, 1938.
\newline \href {https://doi.org/10.2307/2267778} {\path{doi:10.2307/2267778}}.

\bibitem{LucasArnal}
Salvador Lucas and Salvador Arnal.
\newblock Trs.tool.
\newblock Tool available at \url{http://tfmserver.dsic.upv.es:8080/TRS.aspx}.

\bibitem{MeseguerGoguen}
Jos\'{e} Meseguer and Joseph~A. Goguen.
\newblock Initiality, induction, and computability.
\newblock In {\em Algebraic Methods in Semantics ({F}ontainebleau, 1982)},
  pages 459--541. Cambridge Univ. Press, Cambridge, 1985.

\bibitem{Moschovakis}
Yiannis~N. Moschovakis.
\newblock Kleene's amazing {S}econd {R}ecursion {T}heorem.
\newblock {\em Bull. Symb. Log.}, 16(2):189--239, 2010.
\newline \href {https://doi.org/10.2178/bsl/1286889124}
  {\path{doi:10.2178/bsl/1286889124}}.

\bibitem{Moss06}
Lawrence~S. Moss.
\newblock Recursion theorems and self-replication via text register machine
  programs.
\newblock {\em Bull. Eur. Assoc. Theor. Comput. Sci. EATCS}, 89:171--182, 2006.
\newblock URL: \url{https://api.semanticscholar.org/CorpusID:35898839}.

\bibitem{Perkins}
Peter Perkins.
\newblock Unsolvable problems for equational theories.
\newblock {\em Notre Dame J. Formal Logic}, 8:175--185, 1967.

\bibitem{Roberts}
David~Michael Roberts.
\newblock Substructural fixed-point theorems and the diagonal argument: theme
  and variations.
\newblock {\em Compositionality}, 5(8), 2023.
\newblock URL: \url{https://doi.org/10.32408/compositionality-5-8}.

\bibitem{Slattery}
Austin Slattery.
\newblock Solution to homework problem in recursion theory class.
\newblock Unpublished ms, Indiana University, 2022.

\bibitem{Statman}
Rick Statman.
\newblock The word problem for {S}mullyan's lark combinator is decidable.
\newblock {\em J. Symbolic Comput.}, 7(2):103--112, 1989.
\newline \href {https://doi.org/10.1016/S0747-7171(89)80044-6}
  {\path{doi:10.1016/S0747-7171(89)80044-6}}.

\bibitem{SternagelM14}
Thomas Sternagel and Aart Middeldorp.
\newblock Conditional confluence (system description).
\newblock In Gilles Dowek, editor, {\em Rewriting and Typed Lambda Calculi -
  Joint International Conference, {RTA-TLCA} 2014, Held as Part of the Vienna
  Summer of Logic, {VSL} 2014, Vienna, Austria, July 14-17, 2014. Proceedings},
  volume 8560 of {\em Lecture Notes in Computer Science}, pages 456--465.
  Springer, 2014.
\newline \href {https://doi.org/10.1007/978-3-319-08918-8_31}
  {\path{doi:10.1007/978-3-319-08918-8_31}}.

\bibitem{Waldmann}
Johannes Waldmann.
\newblock The combinator {${\bf S}$}.
\newblock {\em Inform. and Comput.}, 159(1-2):2--21, 2000.
\newblock RTA-98 (Tsukuba).
\newline \href {https://doi.org/10.1006/inco.2000.2874}
  {\path{doi:10.1006/inco.2000.2874}}.

\bibitem{yang}
Yafei Yang.
\newblock Solution to homework problem in recursion theory class.
\newblock Unpublished ms., Indiana University, 2022.

\end{thebibliography}

\section*{Appendix A: Details on the termination and confluence}

It is well-known that the rewriting system coming from the $\sss$ and $\kkk$ combinators
is confluent (Church-Rosser) but not terminating.   These central results do not help
with fragments such as the ones in this paper: we know of no general 
results that allow one to infer the confluence of one rewrite system from the confluence
of a larger system, and of course we are after positive results on termination.

All of the termination/confluence results in this paper were shown by running 
programs.  We used  AProVE2023~\cite{AProVE}
in connection with the equational system in Figure~\ref{fig-1}.
AProVE2023 tells us that  termination can be shown with a recursive path ordering (RPO) using
the quasi-precedence $\app > [+,\eee,\ddd]$.
Other programs showed the termination using other general results.
\takeout{
Used ordering:
Recursive path order with status [RPO].
Quasi-Precedence:
app2 > [add2, e, d]

Status:
add2: [1,2]
e: multiset
app2: [1,2]
d: multiset
}
Adding $\uuu$ with the rule
 $(\uuu \app x)\app y \to x\app y$ does not destroy the termination; one uses the 
 RPO  with $\app > \eee > [+,\ddd]$, $\uuu >  [+,\ddd]$.  Adding $\uuu \app x \to x\app \eee$ results
 in a loss of termination, as we saw in Section~\ref{section-large}; the example there was also found by AProVe.
\takeout{
Used ordering:
Recursive path order with status [RPO].
Quasi-Precedence:
app2 > e > [add2, d]
u > [add2, d]

Best result here; termination of 

 add(e,y) -> y
 add(y,e) -> y
 add(add(x,y),z) -> add(x,add(y,z))
 app(add(x,y),z) -> app(y,app(x,z))
 app(app(d,x),y) -> app(x,add(y,x)) 
 app(w,add(x,y)) -> add(app(w,x),app(w,y))
 app(app(w,x),y) -> add(y,x)
 app(w,e) -> e
 app(e,x) -> x
 app(d,e) -> e
 app(app(u,x),y) -> app(x,y)

Used ordering:
Recursive path order with status [RPO].
Quasi-Precedence:
[app2, w] > add2
[app2, w] > [e, d]
}
The confluence results are also obtained by machine.
Here the most informative results come from ConCon~\cite{SternagelM14} and from TRS.tool~\cite{LucasArnal}.
None of our systems have critical pairs, and they are locally confluent.


\takeout{ 
left-linear, right-linear. not collapsing, not duplicating, conservative, orthogonal,
con

For this same TRS, http://tfmserver.dsic.upv.es:8080/TRS.aspx tells us 

The TRS is Left-Linear
The TRS is Right-Linear
The TRS is Linear
The TRS is not Collapsing
The TRS is not Duplicating
The TRS is Conservative
The TRS is not Destructive
The TRS is Orthogonal
The TRS is Almost Orthogonal
The TRS is Weakly Orthogonal
The TRS is Confluent
The TRS is Locally Confluent
}

\takeout{
    
\section*{Appendix B: Proof of Theorem~\ref{theorem-characterize-Slattery}}

We assume throughout that neither $t$ nor $u$ is $\eee$,
for if one of these is $\eee$ then so is the other.

Let $n_t$ be the number of summands of $t$ which are $\ddd$ or $\www$,
let $m_t$ be the number of other summands of $t$.
Let $n_u$  and $m_u$ be defined similarly from $u$.

We also cannot have $m_t = 0$.
For if $m_t = 0$, then $t\app\eee \equiv \eee$.
For the same reason,  we cannot have $m_u = 0$.

\begin{claim}\label{claimzero}
$n_t = 0$ and $n_u = 0$, or more generally when the last summands of $t$ and $u$ are non-constants.
    \end{claim}
    
Suppose suppose towards a contradiction that  $n_t = 0$ and $n_u = 0$.
Then we can write 
\[ 
\begin{array}{lcl}
t &  = & (\www\app x_1) + \cdots + (\www\app x_k) \\
u & = & (\www\app y_1) + \cdots + (\www\app y_\ell)
\end{array}
\]
Applying each to $\eee$ shows that 
\[ 
\begin{array}{lcl}
t &  = &y_1 + \cdots + y_\ell \\
u &  = &x_1 + \cdots + x_k \\
\end{array}
\]
These are normal forms, in view of the omitted
parenthesization.
But then the height of the first summand of $t$ (namely $\www\app x_1$) is larger
than the height of the first summand of $u$ (namely $x_1$); and the same argument
shows the reverse.   So we have a contradiction.

\begin{claim}
$n_t = 1$ and $n_u = 0$, or vice-versa.    
\end{claim}

We assume to begin with that the constant in $t$ is $\www$.
In this case, we can write  
\[ 
\begin{array}{lcl}
t &  = &   (\www\app t_1) + \cdots +  (\www\app t_a) + \www \\
u & = & (\www\app u_1) + \cdots + (\www\app u_m)
\end{array}
\]
(We are using~\ref{claimzero} to be sure that there are no summands after $\www$ in $t$.)
By evaluating $u\app\eee$ and counting summands, we see that 
$m = a +  1$.
By doing the same with $t\app \eee$, we get $a = m$.  
This is a contradiction.

Now if we change the summand $\www$ of $t$ to $\ddd$ and evaluate $t\app \eee$, we get $2a$ summands.
So we have $2a = m$, and from above $m = a + 1$.  Thus, $a = 1$ and $m = 2$.
Thus $t = \www\app t_1 + \ddd$, and $u = \www\app u_1 +  \www\app u_2$.
Evaluating $u@\eee$ shows that $t = u_1 + u_2$.   In particular, $u_2 = \ddd$.
So we have 
$t = \www\app t_1 + \ddd$, and $u = \www\app u_1 +  \www\app \ddd$.
Evaluating $t\app \eee$ shows that $u = \www\app t_1 + t_1$, and so $t_1 = \www\app\ddd$.
Then $u = \www\app(\www\app\ddd) +\www\app \ddd$.
But then $t =\www\app\ddd + \ddd$.
This means that $t_1 = \ddd$, and this is a contradiction.

\begin{claim} 
 $n_t = 1$ and $n_u = 1$.
\end{claim}

Suppose that 
\[ 
\begin{array}{lcl}  t  & =  &  (\www\app t_1) + \cdots +  (\www\app t_a) + c + (\www\app t_{a+1}) + \cdots +  (\www\app t_b)\\
u & =& (\www\app u_1) + \cdots +  (\www\app u_j) +  d + (\www\app u_{j+1}) + \cdots +  (\www\app u_k)
\end{array}
\]
Here and below $0\leq a \leq b$ and $0\leq j \leq k$.
If $c = d = \www$, we get $b = k + 1$, and $k = b+1$.
This means that $b > k$ and $k>b$, and of course this is a contradiction.

\paragraph{Subcase:  $c = \www$ and $d = \ddd$ (or vice-versa)}
If $c = \www$ and $d = \ddd$,
we again get $ b =  k + 1$.  By evaluating $u\app \eee \equiv t$ and counting, 
we get 
$2j + (k-j) = b+1$.  So $j + k = b+1$.
Thus $j + k = k + 2$.   This means that $j = 2$.
We only discuss the case that t that $k = 2$ also.  
Then we would have $b = 3$, and either $a = 0$ or $a=1$ or $a = 2$.
Let's say $a = 2$.
We have
\[ 
\begin{array}{lcl}  t  & =  &  (\www\app t_1) +  (\www\app t_2)   + \www   +  (\www\app t_3) \\
u & =& (\www\app u_1) + (\www\app u_2) +  \ddd
\end{array}
\]
Applying both sides to $\eee$, we get
\[ 
\begin{array}{lcl}  
u & =  &    (\www\app t_1) +  (\www\app t_2)   +  t_3  \\
t & =& (\www\app u_1) + (\www\app u_2) +  u_1 + u_2
\end{array}
\]
Thus, $t_3 = \ddd$, and $u_1 = \www$.  We have
\[ 
\begin{array}{lcl}  
u & =  &    (\www\app t_1) +  (\www\app t_2)   +  \ddd  \\
t & =& (\www\app u_1) + (\www\app u_2) +  \www + u_2
\end{array}
\]
From our original expression $t =  (\www\app t_1) +  (\www\app t_2)   + \www   +  (\www\app t_3) $,
we see that the last summand of $t$ is $\www\app \ddd$.  Thus the second summand is $\www\app(\www\app \ddd)$. 
We have
\[ 
\begin{array}{lcl}  
u & =  &    (\www\app t_1) +  (\www\app t_2)   +  \ddd  \\
t & =& (\www\app u_1) + (\www\app (\www\app \ddd)) +  \www + \www\app \ddd
\end{array}
\]
Now $u\app\eee \equiv  (\www\app t_1) +  (\www\app t_2) + t_1 + t_2$, and so $t_1 = \www$
and $t_2 = \www\app \ddd$.  Then $u_1 = \www$ also.
We get
\[ 
\begin{array}{lcl}  
t & =& (\www\app \www) + (\www\app (\www\app \ddd)) +  \www + \www\app \ddd\\
u & =  &    (\www\app \www) +  (\www\app (\www\app \ddd))   +  \ddd  \\
\end{array}
\]


\bigskip
\paragraph{Subcase:  $c = \ddd$ and $d = \ddd$}
If $c = \ddd$ and $d = \ddd$, we get $a+b = k+1$, and  $j + k = b+1$.
Adding, we get $a + b + j + k = b + k + 2$.  So $a + j  = 2$.
Recall that $a, j\geq 0$.
One subcase at this point is when $a = 2$ and $j = 0$.
So we would have
\[ 
\begin{array}{lcl}  t  & =  &  (\www\app t_1) +  (\www\app t_2) + \ddd + (\www\app t_{3}) + \cdots +  (\www\app t_b)\\
u & =&  \ddd + (\www\app u_{1}) + \cdots +  (\www\app u_k)
\end{array}
\]
This is impossible: the first term of $t\app\eee$ is $\www\app t_1$, and this cannot be $\ddd$.

We turn to the subcase when $ a = 1 =j$.
So we have 
\[ 
\begin{array}{lcl}  t  & =  &   (\www\app t_1) + \ddd + (\www\app t_{a+1}) + \cdots +  (\www\app t_b)\\
u & =&  (\www\app u_1) + \ddd + (\www\app u_{j+1}) + \cdots +  (\www\app u_k)
\end{array}
\]
Then applying both sides to $\eee$ gives
\[ 
\begin{array}{lcl}  u  & =  &   (\www\app t_1) + t_1 +  t_{a+1} + \cdots + t_b\\
t & =&  (\www\app u_1) + u_1 + u_{j+1} + \cdots + u_k
\end{array}
\]
Thus $t_1 = \ddd = u_1$, and we must have no terms after $\ddd$ in both of the equations above.
(The proof of this last point is similar to arguments which we have given above.)
The upshot is that $t = (\www\app\ddd) + \ddd = t$.

\begin{claim}
The first summand of both $t$ and $u$ are non-constants,
and $n_t, n_u \geq 2$.
\end{claim}

Let's say that the first terms of $t$ and $u$ are $\www\app x$ and $\www\app y$, respectively.
Then in the normal form of $t\app \eee$, the first term is $\www\app(\www\app x)$, or perhaps it has even more $\www$'s.
In any case, 
\[ 
\wc(\www\app y) = 
\wc(\www\app(\www\app x)) > \wc(\www\app x).\]
The reason for the first equality is that since the first term of $u$ is $\www\app y$, and $t\app e \equiv u$.
The same is true of $u\app \eee$, and the same analysis shows that $\wc(\www\app x)  > \wc(\www\app y)$.
This is a contradiction.

\begin{claim} The first summands of both $t$ and $u$ are non-constants, $n_t, n_u \geq 1$, 
and $|n_t -n_u| \geq 1$.
\end{claim}

Without loss of generality, $n_t \geq 1 + n_u$. In particular, $n_t > 1$.

Let the first summand of $t$ be $\www\app t_1$, and let the first summand of $u$ be $\www\app u_1$.
The first summand of the normal form of $t\app\eee$ is $\www^{n_t} t_1$.
The first summand of the normal form of $u\app\eee$ is $\www^{n_u}  u_1$.
Thus, $\www\app u_1 = \www^{n_t} t_1$, and $\www\app t_1 = \www^{n_u} u_1$.
Removing the outermost $\www's$, we see that
$u_1 =  \www^{n_t-1} t_1$,
and 
$t_1 =  \www^{n_u-1} u_1$.
By the first of these last equations, $t_1$ is a proper subterm of $u_1$.
By the second (and noting that $n_u -1$ might be $0$), 
$u_1$ is a (not necessarily proper) subterm of $t_1$.
This is a contradiction.

\takeout{
Since $u\app \eee \equiv t$, we see that $u_1 = \www\app t_1$.
But $u \equiv t\app \eee$ and  is $\www\app(\www\app(\www\app t_1))$;
it will have even more applications of $\www$ if $t > 3$.
(The reason for this is that the first
step of the computation takes $\www\app t_1$ and runs this on $\eee$, giving $t_1$.  Then the first constant
gives $\www\app t_1$, then the second would give $\www\app(\www\app t_1)$,
and the third gives $\www\app(\www\app(\www\app t_1))$.)
The first term of $t$ is what we get when we apply the first term of $u$ to $\eee$; this is 
$\www\app(\www\app t_1)$.
Thus
$ \www\app t_1 = \www\app (\www\app t_1)$, and this is obviously a contradiction.
}

\takeout{
Let's say that the first terms of $t$ and $u$ are $\www\app x$ and $\www\app y$, respectively.
Then in the normal form of $t\app \eee$, the first term is $\www\app(\www\app x)$, or perhaps it has even more $\www$'s.
In any case, 
\[ 
\wc(\www\app y) = 
\wc(\www\app(\www\app x)) > \wc(\www\app x).\]
The reason for the first equality is that since the first term of $u$ is $\www\app y$, and $t\app e \equiv u$.
The same is true of $u\app \eee$, and the same analysis shows that $\wc(\www\app x)  > \wc(\www\app y)$.
This is a contradiction.
}

\takeout{
\begin{claim} The first summands of both $t$ and $u$ are constants,
and $n_t, n_u = 1$.
\end{claim}

Suppose that \[ 
\begin{array}{lcl}  t  & =  & c + (\www\app t_1) + \cdots +  (\www\app t_k)\\
u & =&  d + (\www\app u_1) + \cdots +  (\www\app u_\ell)
\end{array}
\]
Here $c$ and $d$ are constants.
Then $t\app \eee \equiv t_1 + \cdots  +t_k$.   Since this is a normal form,
we see that $u =  t_1 + \cdots  +t_k$.  Thus, $u$ has $k$ summands.  
But we also see from the original formulation that $u$ has $\ell + 1$ summands.
Thus, $k = \ell + 1$. 
All of this is symmetric, and so we also see that $\ell = k + 1$.   So we have a contradiction.
}

\takeout{
\begin{claim}
 $n_t = 1$ and $n_u = 2$ (or vice-versa),
 and the first summands of $t$ and $u$ are non-constants.
  
\end{claim}

Let
\[ 
\begin{array}{lcl} t &   = &  (\www\app x_1) + \cdots +  (\www\app x_k) + x'\\
u &  = &  (\www\app y_1) + \cdots +  (\www\app y_\ell) + y' + y''\\
\end{array}
\]
where $x'$, $y'$, and $y''$ are either $\www$ or $\ddd$.  
(We have put these ``constant terms'' at the end, but they might be mixed in with 
the other summands.)
By evaluating $t\app\eee$ and comparing the first terms of the normal forms, we see that 
$\www\app x_1 = \www\app y_1$; thus $x_1 = y_1$.
But evaluating $u\app \eee$ then gives the first term of $t$ to be $\www\app(\www\app x_1)$.
This is a contradiction.
}

\begin{claim}
 $n_t =2$ and $n_u = 0$,
 and the first summand of $t$ is a constant (or vice-versa).
 \end{claim}

 We write $t$ and $u$:
 \[ 
\begin{array}{lcl}  t  & =  & c + (\www\app t_1) + \cdots +  (\www\app t_a) + d + (\www\app t_{a+1}) + \cdots +  (\www\app t_b)\\
u & =& (\www\app u_1) + \cdots +    (\www\app u_k)
\end{array}
\]
Here and below $0\leq a \leq b$ and $0\leq k$.

Calculating $u\app \eee$ gives $k = b + 2$.

We have several subcases:

First, we have the case $c = \www$ and $d = \www$.  This gives $ b= k$.
This is a contradiction.

Second, we have $c= \www$ and $d = \ddd$.
We get $a + b = k$.   So $a + b = b + 2$.  Thus $a = 2$.
The rest of this argument is almost the same as the fourth subcase, given in full below.

Third, we have the case $c = \ddd$ and $d = \www$.
This is exactly as in the first subcase, $c = \www = d$.

Finally, we have $c= \ddd$ and $d = \ddd$.
We get $a + b = k$.  This means that $ a= 2$ and $b = k$.
So we have 
 \[ 
\begin{array}{lcl}  t  & =  & \ddd +  (\www\app t_1) +(\www\app t_2) + \ddd + (\www\app t_3) + \cdots +   (\www\app t_{k-2})\\
u & =& (\www\app u_1) + \cdots +    (\www\app u_k)
\end{array}
\]
Claim~\ref{claimzero}
 shows that there cannot be any summands after the second  $\ddd$ in $t$.
That is, $k-2 = 2$; so $k = 4$.
We have
 \[ 
\begin{array}{lcl}  t  & =  & \ddd +  (\www\app t_1) +(\www\app t_2) + \ddd \\
u & =& (\www\app u_1) +   (\www\app u_2) +  (\www\app u_3) +    (\www\app u_4)
\end{array}
\]
So $u_1 = \ddd = u_4$.
Also, 
$u \equiv t\app\eee\equiv  (\www\app t_1) +   (\www\app t_2) + t_1 +    t_2$.
And so $t_2 = \www\app u_4 = \www \app \ddd$.
Furthermore, equating two expressions for $u$ gives $t_1 = u_1 = \ddd$.
Now we know that $t = \ddd +  (\www\app \ddd) +(\www\app (\www\app\ddd)) + \ddd$.
It follows that $u \equiv t \app \eee =   (\www\app \ddd) +(\www\app (\www\app\ddd)) + \ddd + (\www\app \ddd)$.
In particular, the third summand of $u$ is $\ddd$.
This contradicts with what we have above, where the third summand of $u$ is of the form 
$\www\app u_3$.

\takeout{
So $t_1 = u_1$, $t_2 = u_2$, $t_1 =   (\www\app u_3)$, and $t_2 =  (\www\app u_4)$.
Evaluating $u\app\eee$ gives
$u_1 + u_2 + u_3 +u_4 =  \ddd +  (\www\app t_1) +(\www\app t_2) + \ddd$.
So $u_1 = \ddd = u_4$.
We have
 \[ 
\begin{array}{lcl}  t  & =  & \ddd +  (\www\app \ddd) +(\www\app (\www\app\ddd)) + \ddd \\
u & =& (\www\app \ddd) +   (\www\app u_2) +  (\www\app u_3) +    (\www\app \ddd)
\end{array}
\]
}

\begin{claim}
$n_t = 0$ and $n_u \geq 3$, and the first summand $u$ is a constant  (or vice-versa).
\end{claim}

This case is similar to what we have seen.

\begin{claim}
 $n_t = 0$ and $n_u = 2$,
 and the first summands of $t$ and $u$ are non-constants  (or vice-versa).
 \end{claim}
 Let
\[
\begin{array}{lcl}
t &  = &  (\www\app x_1) + \cdots +  (\www\app x_k)\\
u & = &  (\www\app y_1) + \cdots +  (\www\app y_\ell) + z' + z'\\
\end{array}
\]
where $z'$ and $z''$ are $\www$ or $\ddd$, and these terms can be mixed
with the previous ones. 

The number $k$ must be $\ell + 2$, and by easy counting, we see that 
$\ddd$ must be the third term in $u$, and $\www$ the last term.
In more detail, if $\ddd$ were the fourth or greater term, 
then $u\app\eee$ would add at least six terms, and only two ($\ddd$ and $\www$)
would disappear into $t$.   Similarly, if $\ddd$ were the first or second term in $u$,
then $u\app\eee$ would not gain two terms.   This shows that $\ddd$ is exactly the 
third term.  

We next claim  $\www$ must be the last
term in $u$. If not, let 
$j$ be such that $\www$ is the $j$th term in $u$.
The $(j+1)$st term in $u$ is  $\www\app  z_1$ for some $z_1$.
Let the $(j+1)$st term in $t$ be $\www\app  z_2$.
Then when we apply $t$ and $u$ to $\eee$, we see that
$z_1 = \www\app  z_2$, and $z_2 = \www\app  z_1$.
So $z_1 = \www\app  (\www\app  z_1)$. This is a contradiction.

So the form of $t$ and $u$ is
\[
\begin{array}{lclclclclclclclcl}
t & = & (\www\app x_1) &+& & & &  & & &  \cdots &+& (\www\app x_{k-1})
&+& (\www\app x_{k})    \\
u  & = & (\www\app y_1) &+&  (\www\app y_2) &+& \ddd & + & (\www\app y_3) 
& + & \cdots &+& (\www\app y_{k-2}) &+ &\www \\
\end{array}
\]
Again, $\ddd$ is the third term in $u$, so $\www\app y_1$ and
 $\www\app y_2$ are definitely present in $u$.
We next claim that the terms $ (\www\app y_3)$, $\ldots$, 
$ (\www\app y_{k-2}) $ must be missing from the sum above,
 so that $k = 4$.
If not, consider $\www\app x_{k-1}$ and $\www\app y_{k-2}$.
By applying $t$ and $u$ to $\eee$, we have $x_{k-1} = y_{k-2}$, and 
$\www\app y_{k-2} = \www\app x_{k-1}$.  Thus, 
$ y_{k-2} = \www\app y_{k-2}$, and this is a contradiction.

At this point, we see that $t$ and $u$ are of the following forms:
\[
\begin{array}{lclclclclcl}
t & = & (\www\app x_1) &+&  (\www\app x_2) &+& (\www\app x_3) &+&  (\www\app x_4)\\
u  & = & (\www\app y_1) &+&  (\www\app y_2) &+& \ddd & + & \www \\
\end{array}
\]
Apply $u$ to $\eee$  to see that 
\[
\begin{array}{lclclclclcl}
t & = & (\www\app (\www\app y_1)) & + &  (\www\app (\www\app y_2))
& + &  (\www\app y_1) &+&  (\www\app y_2)\\
u & = & x_1 & + & x_2  & + & x_3 & + & x_4\\
 \end{array}
\] 
From the last terms of $u$ just above,
 we see that $x_3 = \ddd$ and $x_4 = \www$.
From the last terms of $t$, we see that $y_1 = x_3$, and $y_2 = x_4$.
So we have 
\[
\begin{array}{lclclclclcl}
t & = &  (\www\app (\www\app \ddd)) & + &  (\www\app (\www\app \www)) &+& (\www\app \ddd) &+&  (\www\app \www)\\
u  & = & (\www\app \ddd) &+&  (\www\app \www) &+& \ddd & + & \www \\
\end{array}
\]
Thus, $t$ and $u$ are Slattery's twins.
}

\end{document}